\documentclass[twocolumn,preprintnumbers,
  superscriptaddress,
  nofootinbib,
  amsmath,
  amssymb,
  aps,
  prd,
  colorlinks=true,
  linkcolor=blue,
  urlcolor=blue,
  citecolor=blue,
  anchorcolor=blue
]{revtex4-2}

\usepackage{wdmmanuscript}
\usepackage{hyperref}
\usepackage{fontawesome5}
\usepackage[normalem]{ulem}
\usepackage{mathtools}
\usepackage{xcolor}
\usepackage{orcidlink}

\DeclareRobustCommand{\codelink}[1]{\,\href{#1}{\mbox{\scalebox{0.65}{\faGithub}}}}

\begin{document}

\preprint{APS/123-QED}

\title{An explicit and differentiable Wilson--Daubechies--Meyer transform for gravitational-wave data analysis}

\author{Avi Vajpeyi \orcidlink{0000-0002-4146-1132}}
\thanks{GM and AV contributed equally to this work.}
\affiliation{Department of Statistics, University of Auckland, 38 Princes St, Auckland, New Zealand}

\author{Giorgio Mentasti \orcidlink{0000-0003-1115-9220}}
\thanks{GM and AV contributed equally to this work.}
\affiliation{Universit\'e Paris Cit\'e, CNRS, Astroparticule et Cosmologie, F-75013 Paris, France}

\author{Quentin Baghi \orcidlink{0000-0002-1555-9283}}
\affiliation{Universit\'e Paris Cit\'e, CNRS, Astroparticule et Cosmologie, F-75013 Paris, France}

\author{Ollie Burke \orcidlink{0000-0003-2393-209X}}
\affiliation{School of Physics and Astronomy, University of Glasgow, University Avenue, Glasgow, G12 8QQ, UK}
\affiliation{Laboratoire des 2 Infinis - Toulouse (L2IT-IN2P3), Université de Toulouse, CNRS, UPS, F-31062 Toulouse Cedex 9, France}

\author{Lorenzo Speri \orcidlink{0000-0002-5442-7267}}
\affiliation{European Space Agency (ESA), European Space Research and Technology Centre (ESTEC), Keplerlaan 1, 2201 AZ Noordwijk, the Netherlands}
\affiliation{Leiden Observatory, Leiden University, P.O. Box 9513, 2300 RA Leiden, the Netherlands}
\begin{abstract}
The Wilson--Daubechies--Meyer (WDM) time-frequency
transform has been widely used in gravitational-wave astronomy, yet a self-contained, mathematically explicit reference for practitioners remains lacking. This is especially true for those wishing to adopt the transform in modern Python and JAX inference workflows.
We present \package, an open-source Python package implementing the WDM wavelet-packet time-frequency transform, and document its mathematical foundations, statistical properties, and practical implementation for gravitational-wave data analysis.
The package supplies NumPy and JAX backends, both transforms (forward and inverse) validated to floating-point precision, with the JAX backend enabling GPU-accelerated transforms of million-point data streams in tens of milliseconds. As a worked example we verify that the WDM-domain likelihood reproduces frequency-domain posteriors for a resolved LISA galactic binary under a shared stationary noise model, confirming numerical equivalence of the two representations in that controlled setting.
This work paves the way for systematic optimisation of WDM tilings -- a particularly promising direction for the non-stationary noise, stochastic backgrounds, and data gaps anticipated in future detectors -- and for direct comparisons with alternative time–frequency representations needed to meet the challenges of future gravitational-wave data analysis.
\end{abstract}

\maketitle

\section{Introduction}

Time-frequency analysis is widely used in gravitational-wave astronomy. It enables template-free characterisation of source signals~\cite{krolak_1996,Virtuoso2024}, instrumental transients~\cite{robinet2020}, and the joint analysis of astrophysical signals and terrestrial disturbances~\cite{Cornish_2015}. Time-frequency representations also provide compact waveform descriptions and facilitate the modelling of non-stationary detector noise, such as Q-transforms \cite{Chatterji:2004qg} and chirplets \cite{ChassandeMottin:2005bm}.
Such techniques are valuable not because frequency-domain analyses are intrinsically invalid for these problems, but because they become difficult to apply in the presence of non-stationary features such as time-varying noise statistics or data gaps. They have proven particularly useful for tracking the harmonic modes of long-lived signals, for example in the search and characterisation of extreme mass-ratio inspirals~\cite{Gair_2005, Gair:2008ec, Speri_2026} and stellar-origin black hole binaries~\cite{bandopadhyay2026globaltimefrequencysearchstellarmass}, both mHz sources targeted by space-based detectors~\cite{2024arXiv240207571C, Luo:2015uga}. Among the available techniques, discrete wavelet-packet transforms are particularly well suited to gravitational-wave data analysis: they furnish localized, orthonormal basis functions capable of representing transient and chirping signals, and for locally stationary noise they yield an approximately diagonal noise covariance, greatly simplifying likelihood evaluation and spectral estimation \cite{Cornish:2020kdz}.

The Wilson--Daubechies--Meyer (WDM) wavelet-packet basis already has a substantial history in gravitational-wave astronomy.  The fast Wilson--Daubechies transform was developed for transient gravitational-wave analysis by Necula, Klimenko, and Mitselmakher~\cite{Necula_2012}, and WDM representations have been used in coherent WaveBurst and related burst-search contexts~\cite{2023PhRvD.107f2002S, 2021SoftX..1400678D,Anderson:2000yy,Klimenko:2004qh}.
These earlier papers and implementations established the theoretical and algorithmic basis of the transform for excess-power and time-frequency analyses, including its orthonormal structure, controlled spectral leakage, flexible frequency sub-bands, and time-delay filters relevant for sky localization.  Subsequent WDM software and applications by Digman and Cornish further developed fast forward and inverse transforms, WDM likelihoods, and time-frequency analyses for LISA sources and backgrounds~\cite{2023ascl.soft07037D, digman_cornish_bbh_wdm,digman_cornish_tv_galactic_wdm}. Recent work by Pearson and Cornish has also shown how WDM-domain methods can be combined with Bayesian data augmentation to handle data gaps in long-duration gravitational-wave observations~\cite{pearson_cornish_wdm_gaps}.

Cornish developed a complementary WDM-domain programme for likelihood-based gravitational-wave inference and waveform generation. In this formulation, regularly sampled data are mapped to an orthogonal grid of localized time-frequency pixels; slowly evolving signals may be represented compactly as tracks on this grid, and suitable approximations of binary waveforms can be generated directly in the WDM domain~\cite{Cornish:2020odn}, although exact waveform generation directly in the WDM domain is in general non-trivial.
A central statistical motivation is that locally stationary noise, whose spectrum varies slowly across an individual time-frequency pixel, has an approximately diagonal covariance in the WDM domain. This approximate diagonalization, analyzed in detail in Ref.~\cite{cornish_nonstationary_wdm}, provides the WDM-domain analogue of the familiar diagonal Fourier-domain covariance for stationary noise and underlies WDM Whittle-type likelihoods.
While the WDM transform has been carefully developed and implemented across several works~\cite{Cornish_WDM_Transform,Digman_WDMWaveletTransforms,Moore_WDM_GW_wavelets,cWB_software,Farr:2025wdm}, here we place particular emphasis on making the correspondence between the equations and their implementation fully explicit and self-consistent.

The present manuscript does not claim to introduce the WDM transform, to replace these foundational analyses, or to outperform existing pipeline-specific implementations.  Its contribution is instead to make the sampled WDM construction easy to adopt and reproduce in modern Python-based gravitational-wave workflows.  We give a self-contained derivation of the forward and inverse transforms, state the packed $(N_t,N_f+1)$ coefficient layout explicitly, spell out the DC and Nyquist channel treatment, and set the normalization, phase, and indexing conventions used by the implementation. These choices are then validated through exact round-trip reconstruction tests -- applying the forward transform followed by its inverse -- and empirical checks of coefficient decorrelation.
The accompanying \package\ package provides CPU and GPU-capable NumPy/JAX backends, so that the same mathematically explicit transform can be used in conventional analyses and in differentiable inference workflows.

The motivation for the WDM can be understood by contrasting the two standard representations of a discrete signal. In the time domain one can identify when a feature occurs but not which frequency band is active, whereas in the Fourier domain one can identify which frequencies are present but not when they appear. Many gravitational-wave signals and detector artefacts are neither purely stationary nor purely impulsive, so neither representation alone is fully satisfactory.
For what concerns time-frequency representations, Short Fourier Transforms (STFT) have deep history within the Continuous Wave community for long-lived gravitational wave sources suitable for ground-based detectors \cite{Krishnan:2004sv,Palomba_2005}: STFTs often make use of compactly-supported sliding windows such as square or Tukey windows. In contrast, WDM employs atoms whose corresponding time-domain representation exhibits algebraically decaying tails, falling as $1/t^3$ (see Fig.~\ref{fig:architecture}, panels (a) and (b)) for the cosine-tapered window adopted in this work, being well compact in time and frequency (with the added expense of loss of frequency resolution in comparison to the STFT).
This makes the WDM domain an excellent domain of choice for burst like sources, such as massive black holes for LISA while still being suitable for long-duration band-limited signals such as galactic binaries and stellar origin black hole binaries.
However, there is a more fundamental difference between the two representations.
The STFT is not \textit{critically sampled} as defined in Ref.~\cite{Cornish:2020odn}. A mathematically complete representation carries both positive- and negative-frequency indices; for the real-valued time series typically encountered in gravitational-wave data analysis these are related by Hermitian symmetry and need not be stored explicitly, just as the real FFT keeps only $N/2+1$ of the $N$ complex Fourier coefficients. The redundancy of the STFT arises instead from its overlapping, oversampled sliding windows, which carry more coefficients (namely 2N complex values) than there are independent degrees of freedom. Even if keeping only the positive frequency half of the values in the STFT, a total number of $N$ \textit{complex} values has to be used. On the other hand, the WDM transform only needs $N$ real coefficients (the $m>0$ channels) to represent a time series of $N$ samples. Both bases are orthonormal and admit an exact inverse transform, so the original time series can be reconstructed to floating-point precision in either case; only the WDM, however, achieves this without redundancy.
The property of WDM being \textit{critically sampled} allows a faster computation (by a factor of 2, as discussed above) of the transforms, and a cheaper memory demand.
However, this faster representation comes with a price: since the WDM coefficients are real numbers, the phase information of the signals represented is spread through the various WDM time-frequency bins, and it is less immediate to be tracket compared to STFT. For this reason too, WDM representation is, in principle, more suitable for bursts/power-based searches, while STFT looks preferable for long-lasting signals, where the phase brings useful information.
Therefore, for practical purposes, the WDM transform is largely comparable to a STFT but remains a distinct representation of the data. Which representation is preferable is problem-specific, depending on the sources under study and the windows adopted. We illustrate one such controlled comparison -- frequency-domain versus WDM-domain inference for a resolved LISA galactic binary -- in Sec.~\ref{sec:lisa_gb}.

The remainder of this paper is organized as follows. We first fix notation and conventions in Sec.~\ref{sec:conventions}. Section~\ref{sec:wdm} develops the WDM transform, including the Meyer window, the full three-case basis definition covering DC, interior, and Nyquist channels, and the forward and inverse transforms. Section~\ref{sec:stats} derives the statistical properties of the coefficients, the WDM likelihood and inner product, and the conditions under which the coefficients may be treated as approximately independent. Section~\ref{sec:code_package} describes the \package\ package, presents runtime and accuracy benchmarks, and demonstrates the transform on astrophysical signals. We conclude in Section~\ref{sec:conclusions}. 

We have prioritised readability of the main text of the paper over major mathematical details. We refer the interested reader to Appendix~\ref{app:normality_condition} for an explicit proof on the orthonormality condition of the WDM basis components. Appendix~\ref{app:forward_derivation} details the mathematical calculations related to the forward/inverse transforms. Finally, Appendix~\ref{app:stats} discusses the diagonal approximation of the WDM covariance matrix for a wide-sense stationary process -- one of the key features why the WDM domain is suited for inference within gravitational wave astronomy. 

\subsection{Conventions}\label{sec:conventions}
Consider a data stream consisting of $N$ uniformly sampled values of a real quantity $x$ at times $\{t_n\}_{n=0}^{N-1}$ for $t_n = t_0 + n\Delta t$ over a total observation time $T = N\Delta t$:
\begin{align}\label{st_definition}
  x[n]=x(t_n)\qquad,\qquad n=0,\dots,N-1\;,
\end{align}
with sampling interval $\Delta t = t_{n+1} - t_{n}$. All available information about the signal $x[n]$ is encoded in these $N$ samples. We define the sampling rate, the Nyquist frequency and the frequency spacing as
\begin{align} F_s&\equiv\frac{1}{\Delta t}\equiv\frac{N}{T}\;,\nonumber\\
f_{\rm max}&=\frac{1}{2} F_s\;,\nonumber\\
\Delta f &= 1/T\,.
\end{align}
In that regard, if we set $t_0=0$, we can write $t_n=n\Delta t$.
In the frequency domain, the data stream $\{x_n\}_{n=0}^{N-1}$ can be discrete-Fourier-transformed (DFT) to a set of other $N$ complex numbers as
\begin{equation}
  \label{eq:dft-definition}
  \tilde x[l] = \sum_{k = 0}^{N-1} x[k]e^{-2\pi i l k/N},
\end{equation}
for Fourier frequencies
\begin{equation*}
  f_{l} = \{-N\Delta f/2, \ldots, -\Delta f, 0, \Delta f, \ldots, (N/2 - 1)\Delta f\},
\end{equation*}
i.e.\ the zero-frequency bin is centred in the array. All transforms presented in this work are dimensionless and use the standard \texttt{numpy}'s convention. This is in contrast to~\cite{Cornish:2020odn}.

We assume that the output of a gravitational-wave detector, $x(t)$, is expressed as a linear combination of a signal, $h(t; \boldsymbol{\theta})$, determined by a finite set of (unknown) parameters, $\boldsymbol{\theta}$, and instrumental noise, $n(t)$. If we ignore the presence of calibration errors~\cite{Savalle_2022}, the content of a single data stream, i.e., one output channel from one detector, can be written in the time and frequency domain as:
\begin{align}
  {x}[k] & = h[k;\boldsymbol{\theta}] + n[k],\\
  \tilde x[l]& = \tilde{h}[l;\boldsymbol{\theta}]+ \tilde{n}[l]
\end{align}
where the tilde indicates the Fourier transform.

In the following, bold faced quantities $\boldsymbol{A}$ stand for vector/matrix quantities with components $A_{ij}$. Dagger based operations $\boldsymbol{A}^{\dagger}$ stand for the conjugate transpose of the matrix $\boldsymbol{A}$. 

\section{The Wilson--Daubechies--Meyer (WDM) Transform}\label{sec:wdm}

\subsection{Terminology}
The name Wilson--Daubechies--Meyer combines several pieces of time-frequency analysis terminology. A standard discrete wavelet transform usually refers to a multiresolution, dyadic decomposition (``dyadic'' meaning organized by factors of two, so that each successive scale changes the time and frequency resolution by powers of two). Such transforms represent low-frequency structure with broad time support and fine frequency resolution, and high-frequency structure with narrower time support and coarser frequency
resolution.
A \emph{wavelet packet} (or \emph{wavelet-wavepacket}) transform is more general: it allows a broader class of admissible tilings of the time-frequency plane. The object studied in this paper is the Wilson--Daubechies--Meyer (WDM) wavelet-packet transform; for brevity, we refer to it as the WDM transform throughout. Unlike a standard dyadic wavelet transform, the WDM transform used here has a uniform rectangular tiling, so that all pixels have the same time width $\Delta T$ and frequency width $\Delta F$, with $\Delta T\,\Delta F = 1/2$ under the one-sided frequency convention used below.%

It is also worth clarifying how the WDM transform relates to the more familiar wavelet and wavelet packet transforms, with which it is easily conflated \cite{Johnson:2026rrn}. Ordinary wavelet transforms generate their basis by translating and dilating a single mother wavelet, and the resulting dyadic tiling is logarithmically spaced---trading frequency resolution for time resolution at high frequencies and vice versa. The WDM transform instead tiles the plane by translations and frequency modulations of a fixed window, yielding cells of uniform height. Wavelet and wavelet packet transforms are affine constructions, built by dilating and halving the frequency axis, while the WDM transform is a Heisenberg-group construction, modulating one fixed window with sines and cosines. The Meyer scaling function admits both uses---as the WDM window and as the Meyer mother wavelet---but this overlap is only nominal, since the WDM transform performs no dilations at all.

The naming of the transform reflects two distinct ingredients: the
\emph{Wilson}-basis pairing of positive- and negative-frequency atoms,
introduced together with the phase factor, and the smooth
\emph{Daubechies--Meyer} frequency window used to
construct those atoms, all of which are introduced together with their closed forms in
Sec.~\ref{sec:wdm} below~\cite{Daubechies:1991wv}.

\begin{figure*}[tp]
    \centering
    \includegraphics[width=1.0\linewidth]{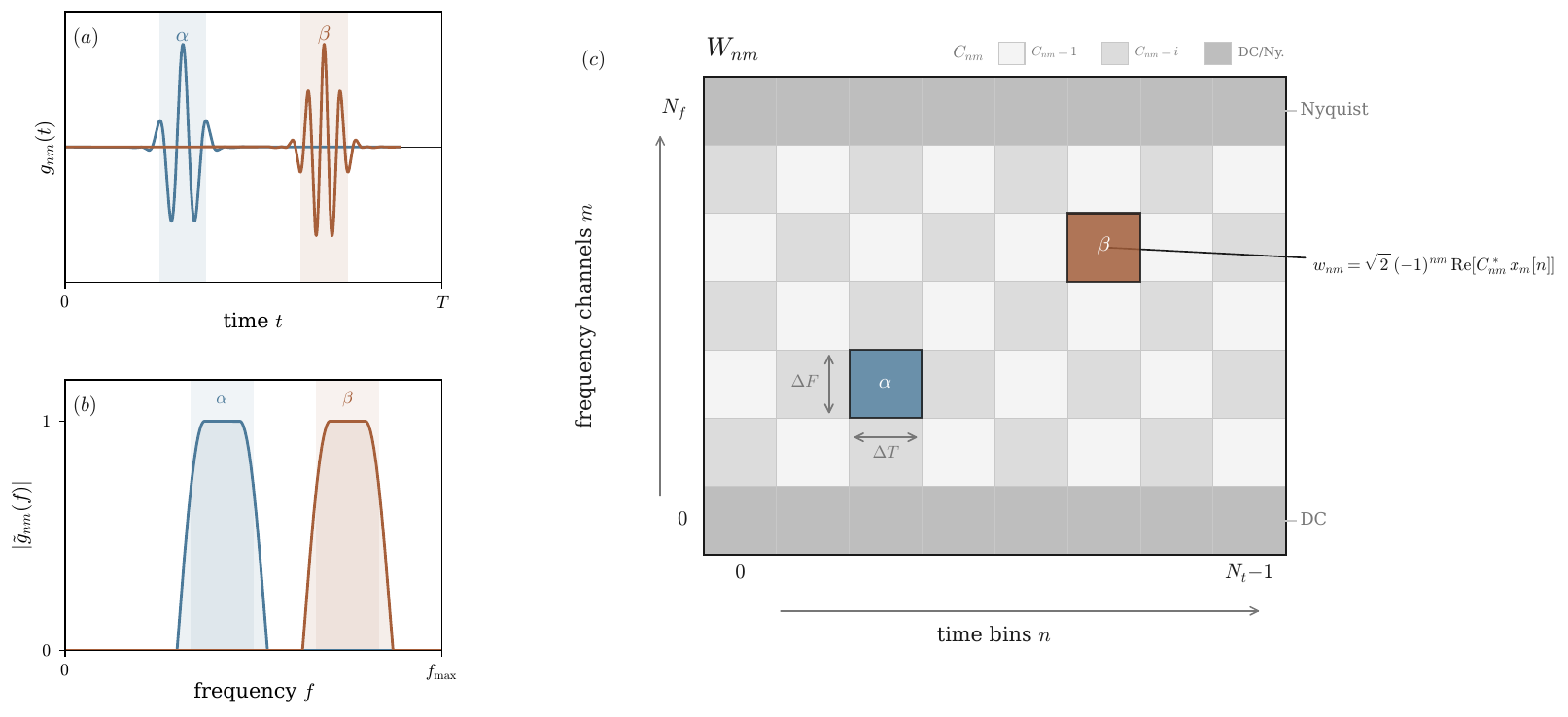}%
    \caption{
        Schematic illustration of the WDM forward transform.
        (a)~Time-domain atoms $g_{nm}(t)$ for two representative WDM cells,
        labelled $\alpha$ and $\beta$. The higher-frequency cell $\beta$ has a
        more rapidly oscillating atom.
        (b)~Frequency-domain magnitudes $|\tilde{g}_{nm}(f)|$ for the same
        two cells, showing their localization around different frequency
        channels.
        (c)~The packed coefficient array $W_{nm}$, with time bins $n$ along
        the horizontal axis and frequency channels $m$ along the vertical axis.
        The highlighted cells correspond to the atoms shown in panels (a) and
        (b). Each interior cell stores
        $w_{nm} = \sqrt{2}\,(-1)^{nm}\,\mathrm{Re}[C^*_{nm}\,\tilde x_m[n]]$
        (Eq.~\protect\eqref{eq:wnm_compact}); the phase factor $C_{nm}$
        alternates between $1$ and $i$ on a checkerboard pattern, while the DC
        and Nyquist edge channels are stored separately.
    }
    \label{fig:wdm_transform}
    \script{wdm_overview_panels.py}
\end{figure*}

The WDM transform maps the original $N$ samples to a packed coefficient grid with $N_t$ time bins and $N_f+1$ stored frequency channels. The transform size is still set by
\begin{align}\label{NtNf_def}
N=N_t\,N_f\;,
\end{align}
and the stored channels are indexed by $m=0,\dots,N_f$, with $m=0$ the DC edge channel, $m=1,\dots,N_f-1$ the interior channels, and $m=N_f$ the Nyquist edge channel. The underlying idea is to partition the data stream into $N_t$ time segments and to perform a frequency domain analysis of each one. Here $N$ is fixed by the data, while $N_t$ is a free user choice subject only to $1 \le N_t \le N$ (and consequently $1 \le N_f \le N$). Given that choice we can define the spacing in both time and frequency with respect to the underlying tiling with area $N = N_{t}N_{f}$
\begin{align}\label{DTDF}
  \Delta T&\equiv\frac{T}{N_t}\; \qquad \text{grid time
  spacing},\nonumber\\
  \Delta F&\equiv\frac{f_{\rm max}}{N_f} \qquad \text{grid frequency spacing}\,.
\end{align}
One immediately sees that
\begin{align}
\label{eq:timefreq-resolution}
  \Delta T &= \frac{T}{N_{t}} = \frac{N\Delta t}{N_{t}}
  = N_{f}\Delta t, \\
  \Delta F &= \frac{f_{\text{max}}}{N_{f}} = \frac{1}{2N_{f}\Delta t}=\frac{N_t\Delta f}{2} = \frac{1}{2\Delta T}.
\end{align}
This final expression shows how the Nyquist theorem still holds. Once you fix $N_t$ (or equivalently $N_f$, $\Delta T$ or $\Delta F$) the grid is uniquely identified. 
Figure~\ref{fig:wdm_transform} summarizes the packed layout, the DC and Nyquist edge channels, the alternating phase factor used for the interior coefficients, and the time- and frequency-domain structure of two representative WDM cells.
Note that $N_t$ and $N_f$ are required to be even — they need not be powers of two. Relaxing the evenness requirement is left to future work.

\subsection{Forward transform}

For a given DFT frequency series $\tilde x[l]$, the WDM transform represents the data through the WDM coefficients $w_{nm}$
\begin{align}\label{wnm_def}
w_{nm}&=\sum_{l=-N/2}^{N/2-1}\tilde x[l]\tilde g^*_{nm}[l]\,,\codelink{https://github.com/pywavelet/wdm_transform/blob/v0.5.0/src/wdm_transform/transforms/xp_backend.py\#L225}
\end{align}
where $n$ corresponds to the time index and $m$ corresponds to the
frequency-channel.

The Wilson-basis elements $\tilde{g}$ with Daubechies-Meyer frequency-domain window $\tilde{\varphi}$ are defined in closed form in the discrete frequency domain below. For DC ($m=0$), interior channels ($0 < m < N_f$), and
Nyquist ($m=N_f$) we have that:
\begin{widetext}
\begin{align}
\label{eq:gnm_full_discrete}
\text{\textbf{Wilson Basis:}}\quad \tilde{g}_{nm}[l] &=
\dfrac{1}{\sqrt{2}}\begin{cases}
e^{-4\pi i n l/N_t}\, \tilde{\varphi}[l], & m=0, \\[6pt]
e^{-2\pi i n l/N_t}\Big(C_{nm}\,\tilde{\varphi}[l - mN_t/2] + C_{nm}^*\,\tilde{\varphi}[l + mN_t/2]\Big), & 0 < m < N_f, \\[6pt]
e^{-4\pi i n l/N_t}\Big(\tilde{\varphi}[l - N/2] + \tilde{\varphi}[l + N/2]\Big), & m = N_f, 
\end{cases}
\codelink{https://github.com/pywavelet/wdm_transform/blob/v0.5.0/src/wdm_transform/windows.py\#L129}\\
\text{\textbf{Daubechies--Meyer:}} \quad  \tilde\varphi[l]&=\sqrt{\frac{2}{N_t}}\begin{cases}
1 & |l_r|<A \\
\cos \left[\frac{\pi}{2}\,\left(
\frac{|l_r|-A}{B}\right)\right] & A \leq|l_r| < A+B\,, \\
0 & \text{otherwise}\,,
\end{cases} \qquad l_r =l\frac{2}{N_t}\,,\codelink{https://github.com/pywavelet/wdm_transform/blob/v0.5.0/src/wdm_transform/windows.py\#L71} \label{eq:DM_window_discrete}\\
  C_{nm}&=e^{\frac{i\pi}{4}(1-(-1)^{n+m})} =
  \begin{cases} 1 & n+m \text{ even,} \\ i & n+m \text{ odd.}\label{eq:c_nm_def}
\end{cases}\codelink{https://github.com/pywavelet/wdm_transform/blob/v0.5.0/src/wdm_transform/windows.py\#L58}
\end{align}
\begin{equation}\label{eq:A_B_condition}
B = 1 - 2A \,, \quad \text{for}\ A\in(0,1/2)\,. 
\end{equation}
\end{widetext}
Equation~\eqref{eq:A_B_condition} gives freedom to the user to define the properties of the Daubechies--Meyer window defined in Eq.~\eqref{eq:DM_window_discrete}.

From this point onwards, we will provide an interpretation for each of the equations \eqref{eq:gnm_full_discrete} -- \eqref{eq:A_B_condition}. We begin by discussing the DC, Interior, Nyquist split of the basis functions $\tilde{g}$, an important feature of our codebase, and the practical and theoretical consequences of the Daubechies--Meyer window $\tilde{\varphi}$.

% This orthogonality is the key reason that the WDM formalism is adept for treating stationary noise processes for reasons that will be discussed later.

From Eq.\eqref{eq:c_nm_def}, the interior channels carry an alternating phase factor $C_{nm}$, while the DC and Nyquist edge channels have a doubled time-shift exponent and no $C_{nm}$ modulation. The phase factor $C_{nm}$ is the
``\emph{Wilson}'' ingredient of the WDM construction: it combines the
positive- and negative-frequency windowed Fourier atoms centered at $mN_t/2$ into a single real, orthogonal pair, which is what
makes the interior basis real-valued and Wilson-orthogonal, to be discussed later in this section, in the
discrete setting. 

These edge channels $m = 0$ and $m = N_f$ in \eqref{eq:gnm_full_discrete} each contribute $N_t$ real degrees of freedom to
the packed grid: the interior block stores $N_t(N_f{-}1)$ real
coefficients via the $C_{nm}$-modulated pairing of the
$\pm mN_t/2$ atoms (the Wilson-basis construction), while the DC
and Nyquist channels each contribute $N_t$ unpaired real coefficients.
The total $N_t(N_f{+}1)$ stored entries therefore represent the same
$N=N_tN_f$ time-domain degrees of freedom, with the edge channels
being exactly $N_t$ real numbers~\cite{Necula_2012,Cornish:2020odn}. 
We refer the reader to Fig.~\ref{fig:wdm_transform} for a schematic of the WDM transform and the packed coefficient layout.  % 

The window $\tilde\varphi$ is the ``\emph{Daubechies--Meyer}''
ingredient of the WDM construction: a smooth, compactly supported
frequency window of the Meyer family, whose taper region overlaps
between adjacent channels in a way that yields exact partition-of-unity
reconstruction. The Meyer family is parametrized by an integer order
$d \geq 1$ that controls the smoothness of the taper: the cosine
argument $(|l_r|-A)/B$ in Eq.~\eqref{eq:DM_window_discrete} is replaced
by the normalized incomplete beta function
\begin{equation}\label{eq:nu_d_def}
  \nu_d(y) = \frac{\int_0^y t^{d-1}(1-t)^{d-1}\,\mathrm{d}t}
                  {\int_0^1 t^{d-1}(1-t)^{d-1}\,\mathrm{d}t}\,,
  \qquad y = \frac{|l_r|-A}{B}\,,
\end{equation}
which satisfies $\nu_d(y) + \nu_d(1-y) = 1$ and steepens the roll-off
as $d$ increases. Setting $d=1$ gives $\nu_1(y)=y$ and recovers the
cosine taper of Eq.~\eqref{eq:DM_window_discrete}. In the present
implementation, and in this work, we use this
$d=1$ cosine-tapered member of the family. We do stress that the parameter $d$ is retained in the Application Programming Interface (API) only for
possible future
generalizations.\footnote{Cornish~\cite{Cornish:2020odn} uses $d=4$,
  which yields a flatter passband and steeper roll-off at the expense
  of a wider time-domain support (the window function must be truncated
  at $\pm q\Delta T$ with $q=16$). The $d=1$ cosine taper used here has
  a gentler roll-off but is more compact in time, requiring a smaller
  $q$. In practice, the choice of $d$ trades spectral leakage
  suppression against temporal compactness of the window; for the
  applications considered in this paper the $d=1$ window provides
  adequate frequency separation while keeping the implementation simple
and the time-domain filter short.} By construction, $\tilde\varphi$ has a limited frequency bandwidth
which is controlled by the parameter $A\in(0,1/2)$ in Eq.\eqref{eq:A_B_condition}. 
The frequency localization of the WDM atoms is illustrated in Fig.~\ref{fig:wdm_transform}(b), where the same two highlighted cells are shown in frequency space.
At each wavelet frequency bin $l$, one sees that the window overlaps with adjacent frequency bins.  

For our implementation we use the default $A =1/3$ and
hence $B = 1/3$, so the window is non-zero for normalised frequencies
$\leq 2/3$. This is a default, not a hard-coded
constraint: $A$ is exposed in \texttt{wdm\_transform} as the
parameter \texttt{a} of the \texttt{WDM} class (with $B=1-2A$ derived
automatically), and any choice with $A\in(0,1/2)$ is admissible.%

A final and essential property of the WDM formalism is the principle of orthogonality. The Wilson basis $\tilde{g}$ defined by Eq.~\eqref{eq:gnm_full_discrete} combined with the Daubechies--Meyer window Eq.~\eqref{eq:DM_window_discrete} with associated constants~\eqref{eq:c_nm_def} provide an orthogonal basis\footnote{Our definitions of the Wilson basis functions \eqref{eq:gnm_full_discrete} carefully include the $m=0$ and $m = N_f$ in order for orthogonality to hold.}, with both of the following equalities proved in Appendix~\ref{app:normality_condition} using Eq.~\eqref{eq:DM_window_discrete} and Eq.~\eqref{eq:c_nm_def}:%
\begin{widetext}
\begin{align}\label{eq:orthonormality}
\sum_{l=-N/2}^{N/2-1}\tilde{g}^*_{nm}[l]\tilde{g}_{pq}[l]&=\delta_{m,q}\begin{cases}
    \frac{1}{2}\left(\delta_{n,p}+\delta_{n,p\pm \frac{N_t}{2}}\right) \qquad m\in\{0,N_f\}\\
    \delta_{n,p} \qquad 0<m<N_f
\end{cases}\nonumber\\
\sum_{n=0}^{N_t-1}\sum_{m=0}^{N_f}\tilde{g}^*_{nm}[l]\tilde{g}_{nm}[l']&=\delta_{ll'}\;,
\end{align}
\end{widetext}

The basis functions here are represented in the frequency domain and it is easily proved they are orthogonal in the time domain.
Orthogonality is one of the key reasons that the WDM formalism is adept for gravitational-wave data analysis, which will be discussed at a later stage of this work.

These are the main ingredients behind the packed WDM grid. A single window shape is shifted across
frequency to define the channel index $m$, and each shifted window is then modulated in time to define the atom indexed by $n$. This is why one WDM coefficient can be read as the strength of one localized time-frequency pattern, rather than as the amplitude of a globally supported sinusoid. 

More concretely, a single coefficient $w_{nm}$ answers the question:
``how much does the data resemble the atom centered at time bin $n$
and frequency bin $m$?'' A large magnitude means that the data
contains a feature with roughly the same time localization and central
frequency. 
Reading the WDM plane coefficient-by-coefficient in this way is often much more
intuitive than interpreting a global Fourier series, because the
index pair $(n,m)$ already carries an approximate time-frequency location.

A further point worth emphasizing concerns how phase information is carried in the two representations \cite{Johnson:2026rrn}. In the Fourier domain the amplitude and phase of a monochromatic signal---its two degrees of freedom---map directly onto the real and imaginary parts of a single complex coefficient, so that a pure cosine is purely real and a pure sine purely imaginary. In the WDM domain the coefficients are instead strictly real, and these same two degrees of freedom are redistributed across a pair of adjacent time cells of opposite parity within a fixed frequency band $m_0$: the amplitude is recovered from the sum of the squared coefficients and the phase from their ratio, $\tan\phi$. The content of a single complex Fourier coefficient is thus spread over two real WDM coefficients. A direct consequence is that the WDM series admits no term-by-term limit to the Fourier series---two real numbers cannot be collapsed into a single complex one---and the two descriptions coincide only in the degenerate case $N_f \to N/2$, $N_t = 2$

In the next subsections, we will provide formulae for the $w_{nm}$ coefficients for both the interior, DC and Nyquist channels separately. In our final subsection, we investigate the orthogonal nature of the Wilson-basis.  

\subsubsection{Interior channels \texorpdfstring{$(0 < m < N_f)$}{(0 < m < N_f)}}

Starting from Eq.~\eqref{wnm_def} with the interior-channel basis
element from Eq.~\eqref{eq:gnm_full_discrete}, the forward transform can be
evaluated in the Fourier domain using
\begin{align}
  w_{nm} = \sum_{l=-N/2}^{N/2-1} \tilde{x}[l]\,\tilde{g}_{nm}^*[l]\,,
\end{align}
For a real-valued input signal, $\tilde{x}[-l] = \tilde{x}^*[l]$, and the real-valued window satisfies $\tilde{\varphi}[-l] = \tilde{\varphi}[l]$.
Splitting the sum into positive and negative frequency halves and
using these symmetries, the negative-frequency contribution is the
complex conjugate of the positive-frequency contribution (see
Appendix~\ref{app:forward_derivation} for details), giving
\begin{equation}\label{eq:wnm_interior}
\begin{split}
  w_{nm} = \sqrt{2}&\,(-1)^{nm}\,\Re\Bigl[C_{nm}^*\\
  &\times\sum_{l'=-N_t/2}^{N_t/2-1}\tilde{x}[l'{+}mN_t/2]\,e^{2\pi i l'n/N_t}\,\tilde{\varphi}[l']\Bigr],\codelink{https://github.com/pywavelet/wdm_transform/blob/v0.5.0/src/wdm_transform/transforms/xp_backend.py\#L92-L102}
\end{split}
\end{equation}
where $l' = l - mN_t/2$ and $\Delta f = 1/T$.
The expression above can be written compactly as
\begin{align}\label{eq:wnm_compact}
  w_{nm} = \sqrt{2}\,(-1)^{nm}\,\Re\,C_{nm}^*\,\tilde x_m[n]\,,
\end{align}
where
\begin{align}\label{eq:xm_def}
  \tilde x_m[n] = \sum_{l=-N_t/2}^{N_t/2-1} e^{2\pi i l n/N_t}\,\tilde x[l +
  mN_t/2]\,\tilde{\varphi}[l],
\end{align}
The quantity
$\tilde x_m[n]$ can be computed using a length-$N_t$ FFT for each channel
$m$, so the total cost of the interior channels is $N_f \times
\mathcal{O}(N_t\log N_t)$.

\subsubsection{DC channel \texorpdfstring{$(m = 0)$}{(m = 0)}}

The DC edge channel has a doubled time-shift exponent and no $C_{nm}$
factor in the basis definition~\eqref{eq:gnm_full_discrete}. Following the
same Fourier-domain evaluation yields
\begin{align}\label{eq:wn0}
  w_{n0} = \sqrt{2\,}\Re\left[\sum_{l=1}^{N_t/2-1}\tilde{x}[l]\,e^{4\pi i l
  n/N_t}\,\tilde{\varphi}[l]\right] + \frac{1}{\sqrt{2}}\tilde{x}[0]\,\tilde{\varphi}[0].\codelink{https://github.com/pywavelet/wdm_transform/blob/v0.5.0/src/wdm_transform/transforms/xp_backend.py\#L80-L90}
\end{align}
Note the factor of $4\pi$ in the exponent (versus $2\pi$ for interior
channels): the DC basis element oscillates at twice the rate in the
time index because it sees both positive and negative frequency
copies of the window centered at $f=0$.

\subsubsection{Nyquist channel \texorpdfstring{$(m = N_f)$}{(m = N_f)}}

The Nyquist edge channel is handled analogously to the DC channel but
centered at the Nyquist frequency $f_{\rm max}$. Its structure
mirrors the DC case with appropriate frequency shifts. In practice,
the Nyquist channel is computed using the same windowed-FFT machinery
as the interior channels, with the frequency shift set to $mN_t/2 = N/2$.
\begin{align}\label{eq:wnNf}
  w_{nN_f} &= \sqrt{2}\,\Re\!\left[\sum_{l=N/2-N_t/2}^{N/2-1}
    \tilde{x}[l]\,e^{4\pi i l n/N_t}\,\tilde{\varphi}[l-N/2]\right]
    \nonumber\\
  &\quad + \frac{1}{\sqrt{2}}\,\tilde{x}[N/2]\,\tilde{\varphi}[0]\,.\codelink{https://github.com/pywavelet/wdm_transform/blob/v0.5.0/src/wdm_transform/transforms/xp_backend.py\#L104-L114}
\end{align}

Collecting the results of each of the $w_{nm}$ coefficients above, the forward transform reads
\begin{widetext}
\begin{align}
\label{eq:forward_transform_discrete}
w_{nm} = \sqrt{2}(-1)^{mn}\Re\,
\begin{cases}
\,\displaystyle\sum_{l=1}^{N_t/2 - 1} \tilde{x}[l]\, \tilde{\varphi}[l]\, e^{4\pi i n l/N_t}\,
+ \dfrac{1}{2}\, \tilde{x}[0]\, \tilde{\varphi}[0],
& m = 0, \\[12pt]
 C_{nm}^{*} \displaystyle\sum_{l' = -N_t/2}^{N_t/2 - 1} \tilde{x}[l' + mN_t/2]\, \tilde{\varphi}[l']\, e^{2\pi i l' n/N_t}\,,
& 1 \le m \le N_f - 1, \\[12pt]
\displaystyle\sum_{l = N/2 - N_t/2}^{N/2 - 1} \tilde{x}[l]\, \tilde{\varphi}[l - N/2]\, e^{4\pi i n l/N_t}
+ \dfrac{1}{2}\, \tilde{x}[N/2]\, \tilde{\varphi}[0],
& m = N_f,
\end{cases}
\codelink{https://github.com/pywavelet/wdm_transform/blob/v0.5.0/src/wdm_transform/transforms/xp_backend.py\#L65-L116}
\end{align}
\end{widetext}
holding true for $N_f \in 2\mathbb{N}$.

\subsection{Backward Transform}

From the orthogonality condition of the Wilson basis, the inverse WDM transform for the Fourier domain $\tilde{x}$ is immediate

\begin{align}\label{eq:inverse_fourier}
  \tilde{x}[l] = \sum_{n=0}^{N_t-1}\sum_{m=0}^{N_f}\tilde{g}_{nm}[l]\,w_{nm}\,,\codelink{https://github.com/pywavelet/wdm_transform/blob/v0.5.0/src/wdm_transform/transforms/xp_backend.py\#L451}
\end{align}
found via direct computation
\begin{align}
\sum_{n=0}^{N_t-1}\sum_{m=0}^{N_f}\tilde{g}_{nm}[l]\,w_{nm} &=\sum_{n=0}^{N_t-1}\sum_{m=0}^{N_f}\tilde{g}_{nm}[l]\,\sum_{l'=0}^{N-1}\tilde x[l']\tilde g^*_{nm}[l']  \\ \nonumber &=\sum_{l'=0}^{N-1}\tilde x[l']\delta_{ll'}=\tilde x[l]\nonumber 
\end{align}
where the first equality uses Eq.~\eqref{wnm_def} and the second using Eq.~\eqref{eq:orthonormality}. 

Specifically, the sum over $n$
produces $N_t\delta_{ll'}$ factors via the discrete orthogonality of the complex exponentials, and the sum over $m$ covers the full frequency axis because adjacent windows satisfy $\tilde{\varphi}^2[l] + \tilde{\varphi}^2[l+N_t/2] = \frac{2}{N_t}$ in their overlap region (proven in Appendix~\ref{app:normalization_Meyer}). Together these two identities ensure exact round-trip reconstruction: forward followed by inverse returns the original data to machine precision, as verified numerically in Section~\ref{sec:code_package}. We derive explicit formulae for the backward transformations in Appendix~\ref{app:inverse_transform}:

\begin{widetext}
\begin{align}
\label{eq:inverse_transform_discrete}
\tilde{x}_{m}[l] = \dfrac{1}{\sqrt{2}}
\begin{cases}
\tilde{\varphi}[l]\, \displaystyle\sum_{n=0}^{N_t - 1} w_{n0}\, e^{-4\pi i n l / N_t},
& m = 0,\ \ 0 \le l < N_t/2, \\[10pt]
\tilde{\varphi}[l - mN_t/2]\, \displaystyle\sum_{n=0}^{N_t - 1} C_{nm}\, w_{nm}\, e^{-2\pi i n l / N_t},
& 1 \le m \le N_f - 1,\ \ (m{-}1)\tfrac{N_t}{2} \le l < (m{+}1)\tfrac{N_t}{2}, \\[10pt]
\tilde{\varphi}[l - N/2]\, \displaystyle\sum_{n=0}^{N_t - 1} w_{nN_f}\, e^{-4\pi i n l / N_t},
& m = N_f,\ \ N/2 - N_t/2 \le l \le N/2,
\end{cases}
\codelink{https://github.com/pywavelet/wdm_transform/blob/v0.5.0/src/wdm_transform/transforms/xp_backend.py\#L119-L176}
\end{align}
\end{widetext}
where the original Fourier domain expression $\tilde{x}[l]$ on the output data is given via
\begin{equation}
\tilde{x}[l] = \sum_{m=0}^{N_f} \tilde{x}_{m}[l]
\end{equation}
Equation~\eqref{eq:inverse_transform_discrete} provides an explicit, closed-form expression for the inverse WDM transform in the sampled, packed convention used throughout this work.
\subsection{Orthogonality \& Forward/Backward Transforms - matrix formalism}

The WDM coefficients $w_{nm}$ are pixels that live on a grid of dimension $(N_t, N_f + 1)$. Now let us define $\alpha(n,m) = n(N_f + 1) + m$ as a function that maps rows and columns of the matrix $w_{nm}$ into a flattened real vector $\boldsymbol{w} \in \mathbb{R}^{N_t(N_f + 1)}$. 

% Let $(n,m)$ stand for  $n$ and column $m$ 
% Now let us define a bijection
% \begin{align}
% \alpha\ : \{n\}_{n = 0}^{N_t - 1} &\times \{m\}_{m=0}^{N_f} \longrightarrow \{0, 1, \ldots, N_t(N_f + 1) - 1\} \\
% \alpha(n,m) &\longmapsto \alpha =\{0, 1, \ldots, n(N_f + 1) +m\}\,,
% \end{align}
% where, for any choice of $\alpha = n(N_f +1) + m$ we have the real vector $\boldsymbol{w} \in \mathbb{R}^{N_t(N_f + 1)}$. In other words, take the WDM coefficients (matrix) and flatten the matrix row by row to create a real vector. 

We can then define the complex rectangular matrices $\boldsymbol{G}\in \mathbb{C}^{N_t(N_f +1) \times N}$ with adjoint $\boldsymbol{G}^{\dagger} \in \mathbb{C}^{N \times N_t(N_f + 1)}$
via the linear maps with action on the frequency domain vector $\tilde{\boldsymbol{x}}\in\mathbb{C}^{N}$ so that $\boldsymbol{w} = \boldsymbol{G}\tilde{\boldsymbol{x}}$ with inverse transform $\tilde{\boldsymbol{x}} = \boldsymbol{G}^{\dagger}\boldsymbol{w}$. In component form, one should interpret $(\boldsymbol{G})_{\alpha,l} = \tilde{g}^{*}_{nm}[l]$, with components of $(\boldsymbol{G})_{\alpha,l}$ here read as row $\alpha \mapsto (n,m)$ evaluated at frequency index $l$. In component form, we can recall the results of \eqref{eq:orthonormality}:
\begin{align}
    \left(\mathbf{G}^\dagger \mathbf{G}\right)_{ll'} 
    &= \delta_{ll'}, \label{eq:orthogonality_likelihood_proof}\\[6pt]
    \left(\mathbf{G}\mathbf{G}^\dagger\right)_{(n,m),(p,q)}
    &= \delta_{m,q}
    \begin{cases}
        \dfrac{1}{2}\left(\delta_{n,p} + \delta_{n,\,p\pm N_t/2}\right), \\[6pt]
        \delta_{n,p}, 
    \end{cases} \label{eq:orthonormality_like_cases}
\end{align}
for the first case in Eq.~\eqref{eq:orthonormality_like_cases} the zeroth $m = 0$ and Nyquist $m = N_f$ bins and second case the interior bins $m \in \{1,\ldots, N_f - 1\}$.
From the orthogonal conditions Eqs. (\ref{eq:orthogonality_likelihood_proof}, \ref{eq:orthonormality_like_cases}), one can see that $\boldsymbol{G}^{\dagger}\boldsymbol{G} = \mathbb{I}_{N} \neq \boldsymbol{G}\boldsymbol{G}^{\dagger}$. In figure \ref{fig:wdm_orthogonality_decorrelation} we plot the two basis contractions $\boldsymbol{G}^{\dagger}\boldsymbol{G}$ and $\boldsymbol{G}\boldsymbol{G}^{\dagger}$ over the full time and frequency grids. Equation \eqref{eq:orthogonality_likelihood_proof} returns the identity, whereas equation \eqref{eq:orthonormality_like_cases} exhibits non-trivial off-diagonal entries due to the inclusion of the zeroth and Nyquist bins.  

As a consequence the forward map $\boldsymbol{w} = \boldsymbol{G}\tilde{\boldsymbol{x}}$ embeds the signal space $\mathbb{C}^N$ into the larger packed space $\mathbb{C}^{N_t(N_f+1)}$, and the left-inverse $\boldsymbol{G}^\dagger$ recovers the signal,
\begin{equation}
    \tilde{\boldsymbol{x}} \xmapsto{\;\boldsymbol{G}\;} \boldsymbol{w} = \boldsymbol{G}\tilde{\boldsymbol{x}} \xmapsto{\;\boldsymbol{G}^\dagger\;} \boldsymbol{G}^\dagger\boldsymbol{G}\tilde{\boldsymbol{x}} = \tilde{\boldsymbol{x}} \,,
\end{equation}
so the forward-then-inverse round-trip is exact for every signal. The reverse round-trip, by contrast, fails for an arbitrary packed array $\boldsymbol{w}$ that is not itself the transform of a signal:
\begin{equation}
    \boldsymbol{w} \xmapsto{\;\boldsymbol{G}^\dagger\;} \boldsymbol{G}^\dagger\boldsymbol{w} \xmapsto{\;\boldsymbol{G}\;} \boldsymbol{G}\boldsymbol{G}^\dagger\boldsymbol{w} \neq \boldsymbol{w} \,,
\end{equation}

In other words, starting from an arbitrary time-frequency representation $\boldsymbol{w}$, it is only possible to reconstruct $\boldsymbol{w}$ through inverse-then-forward transforms only if $\boldsymbol{w}$ comes from a valid WDM transformation $\tilde{\boldsymbol{x}}\mapsto\boldsymbol{w}\mapsto\tilde{\boldsymbol{x}}$\footnote{That is similar to what happens in the frequency domain case: one can always do the round-trip transformation of a real valued signal in time domain $x\to \tilde x\to x$ recovering the original one. The converse is not true: a round-trip transform that starts from a generic data stream $\tilde x$ that does not satisfy the reality condition $\tilde x[-l]=\tilde x^*[l]$, will not lead to the original signal ${\rm fft}[{\rm ifft}[\tilde x]]\neq \tilde x$}. Spurious choices of $\boldsymbol{w}$ would never arise in practice, since WDM-based waveform models would be constructed via time-domain/frequency-domain implementations in the first place. 

On the practical side, the exactness of the \emph{forward-then-inverse} round-trip relies on $\boldsymbol{G}^\dagger\boldsymbol{G} = \mathbb{I}_N$, which in turn requires the DC and Nyquist edge channels to be included and implemented correctly. Omitting or mishandling these channels breaks the completeness relation and introduces non-trivial reconstruction errors in $\tilde{\boldsymbol{x}} \mapsto \boldsymbol{w} \mapsto \tilde{\boldsymbol{x}}$. 

The matrix formalism outlined in this section will prove useful when deriving the WDM-based likelihood from first principles. This is the purpose of the next section. 

\section{Statistics within the WDM formalism}\label{sec:stats}

\subsection{The WDM likelihood}

For a Gaussian frequency-domain noise process characterized by a covariance matrix $\boldsymbol{\tilde \Sigma} = \mathbb{E}[\boldsymbol{\tilde x}\boldsymbol{\tilde x}^\dagger]$,
the log-likelihood of the data $\mathbf{\tilde d}$ given a signal model $\mathbf{\tilde h}$, characterised via residuals $\boldsymbol{\tilde x} = \boldsymbol{\tilde d} - \boldsymbol{\tilde h}$ is \cite{Burke:2025bun}
\begin{equation}\label{eq:frequency_domain_like}
    \log \mathcal{L}(d|\boldsymbol{\theta}) = -\frac{1}{2}\tilde{\boldsymbol{x}}^{\dagger}\tilde{\boldsymbol{\Sigma}}^{-1}\tilde{\boldsymbol{x}} - \frac{1}{2}\ln\det \left(\frac{1}{N}\tilde{\boldsymbol{\Sigma}}\right) - \frac{N}{2}\ln(2\pi)\,.
\end{equation}
If the time-domain noise process is assumed to have a circulant structure (therefore stationary with periodic boundary conditions), the frequency domain noise covariance matrix returns a purely diagonal matrix. In this section, we will perform a similar calculation to ~\cite{Burke:2025bun} and transform the frequency domain likelihood \eqref{eq:frequency_domain_like} into the WDM likelihood used for parameter inference.

The WDM covariance matrix $\boldsymbol{\Sigma}_{w}=\mathbb{E}[\boldsymbol{w}\boldsymbol{w}^{\dagger}]\in\mathbb{R}^{N_t(N_f+1)\times N_t(N_f +1)}$ therefore relates to the one in frequency domain via
\begin{align}
    \tilde{\boldsymbol{\Sigma}} &= \mathbb{E}[\tilde{\boldsymbol{x}}\tilde{\boldsymbol{x}}^{\dagger}] = \boldsymbol{G}^{\dagger}\mathbb{E}[\boldsymbol{w}\boldsymbol{w}^{\dagger}]\boldsymbol{G} = \boldsymbol{G}^{\dagger}\boldsymbol{\Sigma}_{w}\boldsymbol{G} \label{def:cov_freq_matrix}\\
    \boldsymbol{\Sigma}_w &= \mathbb{E}[\boldsymbol{w}\boldsymbol{w}^{\dagger}] = \boldsymbol{G}\mathbb{E}[\tilde{\boldsymbol{x}}\tilde{\boldsymbol{x}}^{\dagger}]\boldsymbol{G}^{\dagger} = \boldsymbol{G}\tilde{\boldsymbol{\Sigma}}\boldsymbol{G}^\dagger \label{def:cov_wdm_matrix}
\end{align}
for $\boldsymbol{\Sigma}_{w}\in\mathbb{R}^{N_t(N_f+1)\times N_t(N_f +1)}$ the \emph{real valued} WDM noise covariance matrix. 

Now, for stationary and circulant noise the frequency domain covariance matrix $\tilde{\Sigma}$ is trivially invertible. However, we have that $N = \text{Rank}(\boldsymbol{G}^{\dagger}\boldsymbol{G}) = \text{Rank}(\boldsymbol{G}) = \text{Rank}(\boldsymbol{G}^{\dagger})$, and that $\text{Rank}(\boldsymbol{\Sigma}_w)  \leq N < N_t(N_f+1)$. The matrix $\boldsymbol{\Sigma}_{w}$ is therefore rank-deficient, and we deduce that $\boldsymbol{\Sigma}_{w}$ is not invertible over the full span of pixels $n \in \{0,\ldots, N_t -1\}$ and $m\in\{0, \ldots, N_f\}$.

To make progress, we define the Moore-Penrose pseudo inverse $\boldsymbol{\Sigma}^+_{w}$ of the matrix $\boldsymbol{\Sigma}_w$
\begin{equation}\label{eq:pseudo_inv_cov}
\boldsymbol{\Sigma}_w^{+} = \boldsymbol{G}\tilde{\boldsymbol{\Sigma}}^{-1}\boldsymbol{G}^{\dagger}\,,
\end{equation}
and we can build the WDM likelihood from \eqref{eq:frequency_domain_like}
\begin{align}\label{eq:likelihood_G_Sigma_w}
    \ln\mathcal{L}(w|\boldsymbol{\theta}) &= -\frac{1}{2}\boldsymbol{w}^{T}\boldsymbol{\Sigma}_{w}^{+}\boldsymbol{w}  \\
    & \quad - \frac{1}{2}\ln\det \left(\frac{1}{N_fN_t}\boldsymbol{G}^{\dagger}\boldsymbol{\Sigma}_{w}\boldsymbol{G}\right) - \frac{N_fN_t}{2}\ln(2\pi)\,.
\end{align}
The explicit expressions for the covariance matrix $\Sigma_w$ and its pseudo-inverse are tricky to calculate in practice when including the zeroth and Nyquist bins. Instead, if one neglects these contributions and restricts to the internal frequency bins $m \in \{1, \ldots, N_f-1\}$ then  $\boldsymbol{G}\boldsymbol{G}^{\dagger} = \mathbb{I}_{N_t(N_f -1)}$ and $\boldsymbol{G}^{\dagger}\boldsymbol{G} = \hat{\boldsymbol{P}}$, for $\hat{\boldsymbol{P}}$ the projection matrix that annihilates the DC and Nyquist frequency band through matrix operation. This then implies that the WDM covariance matrix is now invertible. 

Assuming that the noise process is wide-sense stationary, we show in appendix \ref{app:stats} that, as a first-order approximation, the covariance matrix reads
\begin{align}\label{eq:Sigma_W_diagonal}
(\boldsymbol{\Sigma}_w)_{(n,m),(p,q)} &= \frac{N}{2\Delta t} S[mN_t/2]\delta_{np}\delta_{mq}\,, \\
& = \delta_{np}\delta_{mq}S_{nm}\,,\codelink{https://github.com/pywavelet/wdm_transform/blob/v0.5.0/src/wdm_transform/signal_processing.py\#L117-L172} 
\end{align}
for $0 < m < N_f$, where we define the per-pixel spectral level $S_{nm}~\equiv~\frac{N}{2\Delta t} S[mN_t/2]$. For stationary noise, $S_{nm}$ carries no dependence on the time index $n$; for locally stationary noise it generalizes to the evolutionary spectrum evaluated at the pixel centre, $S_{nm} \propto S[n, mN_t/2]$, in agreement with \cite{cornish_nonstationary_wdm}.

The log-likelihood therefore takes the simple form over the interior bins $m \in \{1, \ldots, N_f - 1\}$

\begin{equation}
  \log \mathcal{L}(\boldsymbol{\theta})
  = -\frac{1}{2}\sum_{\substack{n \\ m\in\mathrm{int}}}
  \left[
    \ln(S_{nm})
    + \frac{\bigl(d_{nm}-w_{nm}\bigr)^2}{S_{nm}}
  \right],
\end{equation}
similarly agreeing with \cite{cornish_nonstationary_wdm}. This pixel-diagonal, log-spectrum-plus-residual form is the WDM-domain analogue of the frequency-domain Whittle likelihood, and we refer to it as the WDM Whittle likelihood throughout. We remark here that if one were performing signal inference assuming a known noise model $S_{nm}$, then the first term is a noise parameter $\boldsymbol{\theta}$ dependent correction and can therefore be neglected.

\subsection{The WDM Inner Product}
In a similar way to a frequency domain analysis, under the assumptions in Appendix \ref{app:stats} we obtain the approximated expression for the noise-weighted inner product in the WDM domain:
\begin{align}
    (a|b)_{\mathrm{WDM}} &= \boldsymbol{w_a}^T(\boldsymbol{\Sigma}_w^{\rm int})^{-1}\boldsymbol{w_b} \\
    & = \sum_{\substack{n \\ m\in\mathrm{int}}} \frac{w_{a,nm}\,w_{b,nm}} {S_{nm}}\,,\label{eq:wdm_inner}\codelink{https://github.com/pywavelet/wdm_transform/blob/v0.5.0/src/wdm_transform/signal_processing.py\#L79-L101}
\end{align}
which reduces to the frequency domain inner product via matrix algebra
\begin{align*}
    (a|b)_{\mathrm{WDM}} &= \boldsymbol{w_a}^T(\boldsymbol{\Sigma}^{\rm int}_w)^{-1}\boldsymbol{w_b}  \\ 
    &= \boldsymbol{\tilde{x}_a}^{\dagger}\boldsymbol{G}^{\dagger}\boldsymbol{G}(\tilde{\boldsymbol{\Sigma}})^{-1}\boldsymbol{G}^{\dagger}\boldsymbol{G}\boldsymbol{\tilde x_b} \\
    & = \boldsymbol{\tilde x_a}^{\dagger}\hat{\boldsymbol{P}}(\tilde{\boldsymbol{\Sigma}})^{-1}\hat{\boldsymbol{P}}\tilde{\boldsymbol{x_b}} \\
    & \approx  \boldsymbol{\tilde x_a}^{\dagger}(\tilde{\boldsymbol{\Sigma}})^{-1}\hat{\boldsymbol{P}}\tilde{\boldsymbol{x_b}}\\
    & = \frac{4\Delta t}{N} \text{Re}\left(\sum_{\substack{l> 0 \\ l\,\in\,\mathrm{interior}}}\frac{\tilde{x_a}^{*}[l]\tilde{x_b}[l]}{S[l]}\right) \\
    & \approx \frac{4\Delta t}{N} \text{Re}\left(\sum_{l=1}^{N/2 - 1}\frac{\tilde{x_a}^{*}[l]\tilde{x_b}[l]}{S[l]}\right) \\
    & = (a|b)_{\text{freq}}\,.
\end{align*}
Where each vector and Wilson basis matrix $\boldsymbol{G}$ only contains full time information $n\in \{0,\ldots,N_t-1\}$ but, as a first working approximation, we exclude the zeroth/Nyquist bin $m\in\{1,\ldots,N_f -1\}$\,. The first approximation features through the local-stationary assumption where we assume that $\tilde{\boldsymbol{\Sigma}}$ is diagonal, (allowing $\boldsymbol{\Sigma}$ and $\hat{\boldsymbol{P}}$ commute) and utilising the fact that $\hat{\boldsymbol{P}}^2 = \hat{\boldsymbol{P}}$. The second approximation comes from extending the sum to include the DC components, assuming that the spectral content of GW sources is minimal at the DC and Nyquist frequency (with respect to the noise power, $S$). 

\begin{figure*}[tp]
  \centering
  \includegraphics[width=1\linewidth]{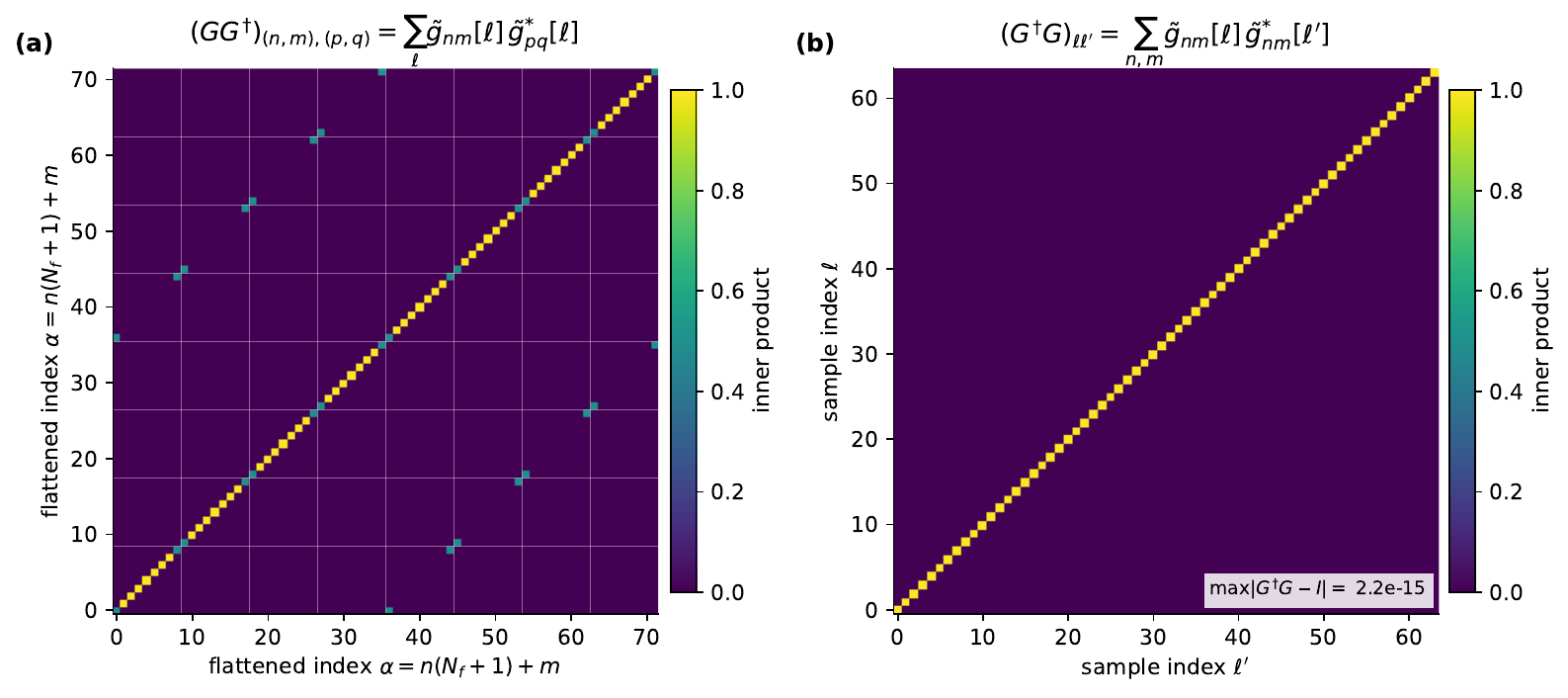}
  \caption{
  \textbf{Orthogonality structure and empirical decorrelation of the
sampled WDM transform.} 
Both panels use the small tiling $N_t = N_f = 8$, $N = N_t N_f = 64$, with packed atoms flattened by $\alpha = n(N_f + 1) + m$. Panel (a) lives in the $N_t(N_f+1) = 72$-dimensional packed coefficient space, while panel (b) lives in the $N$-dimensional signal space. The asymmetry $\boldsymbol{G}\boldsymbol{G}^\dagger \in \mathbb{C}^{72\times 72}$ versus $\boldsymbol{G}^\dagger \boldsymbol{G} \in \mathbb{C}^{64\times 64}$ is intrinsic to the rectangular embedding $G : \mathbb{C}^N \to \mathbb{C}^{N_t(N_f+1)}$ and is the matrix-level expression of the fact that $\boldsymbol{G}^\dagger \boldsymbol{G} = \mathbb{I}_N \neq \boldsymbol{G}\boldsymbol{G}^\dagger$.
Faint grid lines in the panels mark the block boundaries $\alpha = n(N_f+1)$ for $n = 0, 1, \ldots, N_t$, separating successive time slices and making the $\pm N_t/2$ offset structure easy to read off by eye. \textbf{(a)} The packed-space Gram matrix $(\boldsymbol{G}\boldsymbol{G}^\dagger)_{(n,m),(p,q)} = \sum_\ell \tilde g_{nm}[\ell]\,\tilde g^*_{pq}[\ell]$. Interior channels ($m = 1,\ldots,N_f-1$) form an identity block, while the DC and Nyquist edge channels ($m \in \{0, N_f\}$) carry the off-diagonal entries at $p = n \pm N_t/2$ predicted by the first case of Eq.~\eqref{eq:orthonormality_like_cases}. These off-diagonals span the $N_t$-dimensional redundant subspace by which $GG^\dagger$ fails to be the identity.
\textbf{(b)} The signal-space product $(\boldsymbol{G}^\dagger \boldsymbol{G})_{\ell\ell'} = \sum_{n,m} \tilde g_{nm}[\ell]\,\tilde g^*_{nm}[\ell']$, indexed by sample indices $(\ell,\ell')$ rather than the packed pair, equals $\mathbb{I}_N$ to machine precision ($\max|\boldsymbol{G}^\dagger \boldsymbol{G} - \mathbb{I}| \sim 10^{-15}$), demonstrating that the forward-then-inverse round trip is exact.
The exact $\pm 1$ entries at the same $\pm N_t/2$ offsets seen in panel (a) are the empirical signature of the edge-channel redundancy: $w_{n,0}$ and $w_{n + N_t/2,\,0}$ (and likewise for $m = N_f$) are linearly dependent copies of the same underlying real degree of freedom, so their Pearson correlation is identically $\pm 1$ for any input.
}
  \label{fig:wdm_orthogonality_decorrelation}
  \script{wdm_orthogonality_decorrelation.py}
\end{figure*}

These expressions assume that the diagonal approximation~\eqref{eq:Sigma_W_diagonal} holds. When it does not --- for example, when the noise PSD varies rapidly compared to $\Delta T$ --- the off-diagonal covariance terms must be retained and the simple pixel-wise sums above become inadequate.
In practice, if the off-diagonal entries of the empirical covariance remain small after whitening, the diagonal likelihood~\eqref{eq:likelihood_G_Sigma_w} is a safe approximation. The exact Gram-matrix structure shown in Figure~\ref{fig:wdm_orthogonality_decorrelation} underlies this distinction.  A natural question comes to light -- when does the diagonal approximation hold? 

The WDM tiling obeys $\Delta T\,\Delta F = 1/2$ (see Eq.~\eqref{DTDF}), so each atom occupies the minimum allowed time-frequency area consistent with the uncertainty principle, though the actual time-bandwidth product of the cosine-tapered Meyer window exceeds this minimum. The diagonal covariance~\eqref{eq:Sigma_W_diagonal} holds whenever the noise PSD varies slowly compared to this joint support. Concretely, this is the condition of local stationarity, as explained in appendix \ref{app:stats}: spectral variations on timescales shorter than $\Delta T$, or spectral features narrower than $\Delta F$, will induce off-diagonal covariance terms.
The WDM tiling therefore poses an inherent trade-off: choosing finer frequency resolution (larger $\Delta T$) suppresses spectral leakage within each atom but degrades temporal tracking of non-stationarity. In practice a single choice of $N_f$ often suffices for mildly non-stationary noise, but data with sharp spectral features or rapid temporal variation may require a multi-resolution strategy, as discussed by Cornish~\cite{Cornish:2020odn}. Beyond the tiling itself, the shape of the Meyer window -- through the taper width $A$ and taper order $d$ introduced in Sec.~\ref{sec:wdm} -- sets the time-frequency localization of each atom and hence the same leakage--tracking trade-off. Tuning the window is therefore complementary to tuning $N_f$. While in principle many window choices are admissible~\cite{Necula_2012}, in practice the window and its parameters ($A$ and $d$ for the Meyer window) may be optimal for different classes of signals one wishes to represent.

\section{\package\ Package}
\label{sec:code_package}

\begin{figure*}[tp]
  \centering
  \includegraphics[width=\linewidth]{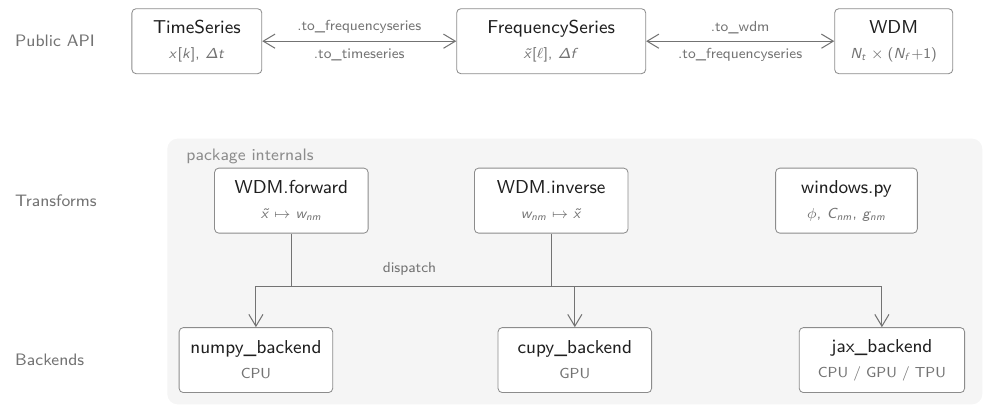}
  \caption{\textbf{Architecture of the \package\ package.} The public API
    exposes the \texttt{TimeSeries}, \texttt{FrequencySeries}, and \texttt{WDM}
    datatypes with reversible conversions between them. The forward and inverse
    WDM transforms, together with the shared window machinery
    (\texttt{windows.py}), are dispatched to interchangeable NumPy, CuPy, and
    JAX backends, so the same high-level code runs on CPU, GPU, or TPU.}
  \label{fig:architecture}
  \script{architecture.tex}
\end{figure*}

The \package\ package provides NumPy, CuPy, and JAX backends for the Wilson--Daubechies--Meyer (WDM) transform, all implementing the packed $(N_t, N_f+1)$ coefficient layout and forward/inverse conventions described above. Figure~\ref{fig:architecture} gives a high-level map of the package: a small set of public datatypes (\texttt{TimeSeries}, \texttt{FrequencySeries}, \texttt{WDM}) with reversible conversions, a transform layer, and a backend-dispatch layer that selects the numerical kernel.  The primary interface exposes the full forward and inverse transforms; in addition, band-limited routines allow the transform to be evaluated over a specified subset of frequency channels, reducing compute cost when only a portion of the time-frequency plane is needed.

From the implementation point of view, the dominant operations in the forward transform are one length-$N$ FFT of the input data together with one length-$N_t$ sub-band transform for each stored channel. This gives a cost of
\begin{align}
  T_{\rm forward} = \mathcal{O}(N \log N) + \mathcal{O}(N \log N_t),
\end{align}
which is asymptotically $\mathcal{O}(N \log N)$ because $N=N_tN_f$ and $N_t \leq N$. The inverse transform has the same scaling up to constant factors: it reconstructs the full spectrum from the packed channels and then applies one inverse FFT back to the time domain. In both directions, the stored coefficient grid contains $\mathcal{O}(N)$ real numbers, and the temporary working arrays used by the implementation are also linear in $N$, so the overall memory footprint is $\mathcal{O}(N)$.

We validate the implementation using round-trip reconstruction tests. For an input time series $\boldsymbol{x}$, we compute its packed WDM coefficients, apply the inverse transform, and compare the reconstructed series $\hat{\boldsymbol{x}}$ with the original. Across the benchmarked problem sizes and backends, the relative reconstruction error
\begin{align}
  \epsilon_{\rm rt} = \frac{\|\hat{\boldsymbol{x}}-\boldsymbol{x}\|_2}{\|\boldsymbol{x}\|_2}
\end{align}
remains at the level of double-precision floating-point roundoff, $\epsilon_{\rm rt} \lesssim 10^{-16}$, providing a direct numerical check that the packed layout, including the DC and Nyquist edge channels, is implemented consistently across both backends.

\begin{figure}[!tb]
  \centering
  \includegraphics[width=1\linewidth]{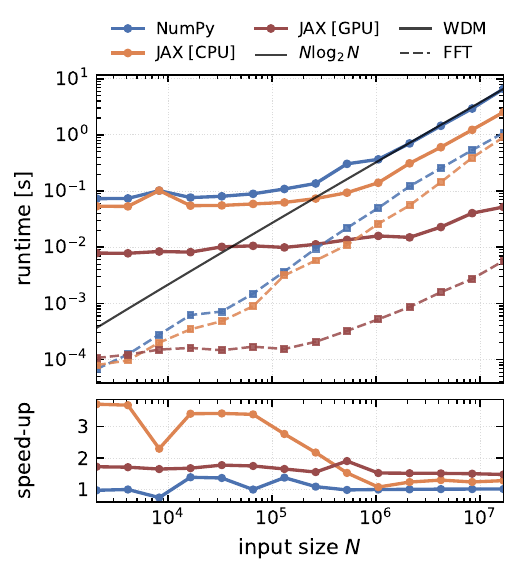}
  \caption{\textbf{Forward WDM runtime versus input length.}
    All transforms use a fixed tiling with $N_t=1024$ time bins and $N_f=N/N_t$ frequency channels.
    \emph{Top}: median single-transform wall-clock runtime over 7 runs
    after one warmup call.  The thin solid black line shows the
    $N\log_2 N$ reference scaling.
    \emph{Bottom}: batched-execution speedup (WDM transform only)
    $t_\mathrm{serial}/t_\mathrm{batch}$ for a batch size of $B=3$,
    where the serial baseline applies the kernel independently to each
    of the three channels and the batched call processes all three
    simultaneously.  Reported timings exclude JAX JIT compilation and
    host--device data transfer.
    }
  \label{fig:runtime_forward}
  \script{benchmark_runtime_snapshot.py}
\end{figure}

All timings are drawn from the benchmark suite included in the package documentation. Each data point reports the median wall-clock time over 7 repeated runs in 64-bit floating-point arithmetic; JAX JIT compilation is excluded via a single warmup evaluation, and GPU timings assume arrays already resident on the device (host--device transfer is not included). Benchmarks were performed on Google Colab using a CPU node (2$\times$ Intel Xeon vCPU, 2.0--2.2\,GHz, 12.7\,GB RAM) and a GPU node (NVIDIA T4, 16\,GB VRAM).

Figure~\ref{fig:runtime_forward} shows the forward-kernel runtime as a function of input size $N$ for the NumPy (CPU), JAX (CPU), and JAX (GPU) backends, together with the batched-execution speedup $t_\mathrm{serial}/t_\mathrm{batch}$. The sweep fixes the number of time bins at $N_t=1024$ and varies $N_f=N/N_t$, so that increasing $N$ refines the frequency resolution at fixed time resolution. We use a batch size of $B=3$, matching the number of time-delay-interferometry data streams (e.g.\ the $A$, $E$, $T$ channels) transformed per likelihood evaluation in a LISA analysis. Here, $t_\mathrm{serial}$ is the wall-clock time for $B$ independent single-channel calls and $t_\mathrm{batch}$ is the time for a single call that processes all $B$ channels simultaneously; larger batches (e.g.\ template banks or ensemble samplers) would benefit further until the device saturates.

At large $N$ all three backends approach $\mathcal{O}(N\log N)$ scaling, consistent with the dominant length-$N$ FFT in the forward transform.  All backends are nearly flat at small $N$: with $N_t$ fixed at $1024$, each call carries a constant setup cost (constructing the length-$N_t$ Meyer window and the per-channel phase factors) that is independent of $N$ and dominates until the $\mathcal{O}(N\log N)$ FFT overtakes it near $N\approx10^5$. This floor sits higher for the CPU backends (per-call Python and memory-allocation costs) than for the GPU. Beyond the threshold GPU efficiency improves and, at $N = 10^6$, the GPU scalar runtime (${\approx}\,16\,\mathrm{ms}$) is roughly $23\times$ lower than NumPy (${\approx}\,370\,\mathrm{ms}$).

The speedup panel shows that batched execution yields no measurable benefit for NumPy ($t_\mathrm{serial}/t_\mathrm{batch} \approx 1$ throughout), as NumPy does not exploit the extra batch dimension for additional parallelism.  JAX (CPU) recovers up to ${\sim}\,3.7\times$ at small $N$ via JIT-compiled vectorisation over the batch axis, with the gain declining toward unity at the largest $N$ as each transform saturates the available cores.
JAX (GPU) yields a roughly constant ${\sim}\,1.5$--$1.8\times$ batched speedup across the full range of $N$, reflecting the three channels sharing the device pipeline.  The $\epsilon_{\rm rt} \lesssim 10^{-16}$ round-trip errors established above hold for all backends and sizes; the runtime differences therefore reflect implementation efficiency rather than any loss of numerical fidelity.

These benchmarks characterise the present NumPy and JAX implementations under the conventions used in this paper, rather than providing an exhaustive comparison with existing pipeline-specific WDM codes. The main conclusion is that the JAX backend can substantially reduce the cost of large WDM transforms while preserving the same floating-point reconstruction accuracy as the NumPy implementation.

The \package\ package enables both a straightforward NumPy-based development path and high-performance accelerator execution via JAX. At small data lengths the runtimes are dominated by per-call overhead rather than arithmetic, so the ranking between backends is sensitive to the platform and measurement methodology. The GPU advantage grows with problem size: once $N$ exceeds $\sim 2^{20}$ the GPU-accelerated backend is markedly faster, while all backends maintain round-trip reconstruction errors at floating-point roundoff.
This combination of flexibility, accuracy, and performance makes \package\ a useful tool for WDM time-frequency analysis of long-duration gravitational-wave signals.

\subsection{Resolved galactic-binary inference with LISA: frequency-domain versus WDM-domain likelihoods}
\label{sec:lisa_gb}

As an end-to-end validation of the WDM-domain likelihood, we compare Bayesian parameter estimation for 100 independent simulations of individual resolved galactic binaries (GBs), with each source analysed in both the frequency domain and the WDM domain with LISA. The two analyses use the same simulated data, source model, priors, and stationary instrumental-noise PSD. The only intended difference is the representation in which the likelihood is evaluated, so the comparison isolates the effect of the frequency-domain to WDM-domain basis change. 
We find that the WDM-domain likelihood reproduces the frequency-domain posterior to numerical precision.

\begin{figure}[!htbp]
  \centering
  \includegraphics[width=\linewidth]{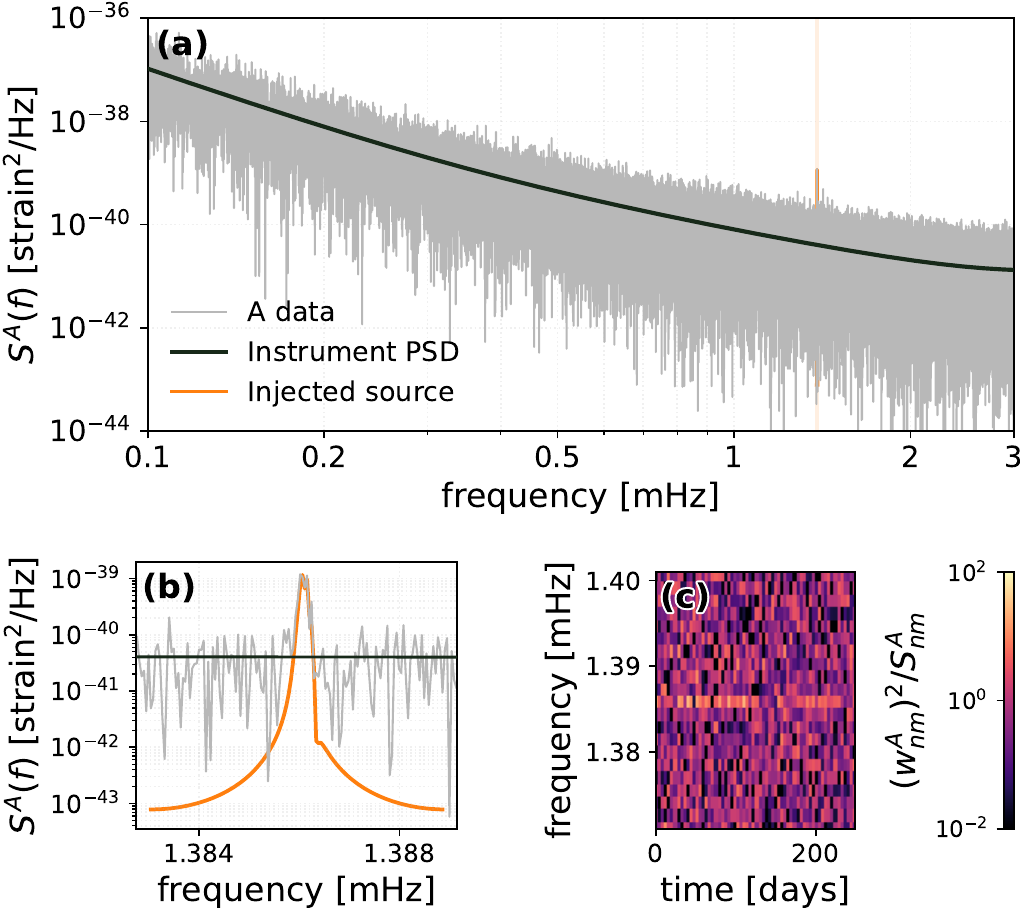}
  \caption{\textbf{Injected resolved galactic binary in both frequency and WDM domains.}
    (a): channel-A frequency-domain data, the instrumental noise PSD, and the
    injected source, with the analysis band shaded. Bottom: zooms of the source
    band in the frequency domain (left (b)) and in the whitened WDM time-frequency
    plane (right (c)), where amplitude is shown in units of the per-pixel noise
    standard deviation.}
  \label{fig:lisa_gb_data}
  \script{lisa_gb_data.py}
\end{figure}

\paragraph{Data and source model.}
The analysis grid spans $(10^{-4},\,3\times10^{-3})\,\mathrm{Hz}$ with cadence $\Delta t = 166.7\,\mathrm{s}$. For each TDI~1.5 A,E,T channel we simulate approximately 8 months ($T\approx253\,\mathrm{d}$, $N = 131072$ samples) of independent stationary instrumental noise, using the PSD model of Ref.~\cite{LISACosmologyWorkingGroup:2022kbp}, and inject a galactic binary generated with \texttt{jaxgb}~\cite{jaxgb} using equal-arm LISA orbits. A representative realization of the channel-A data is shown in Figure~\ref{fig:lisa_gb_data}. The injected source parameters are $f_0 = 1.38599\,\mathrm{mHz}$, $\dot f = 9.39\times10^{-15}\,\mathrm{Hz\,s^{-1}}$, $A = 5.35\times10^{-24}$, $\phi_0 = 1.74\,\mathrm{rad}$, with combined A+E+T optimal SNR$=20.7$. The sky position and orientation are held fixed at their injected values throughout the inference: ecliptic longitude $\lambda = 4.50\,\mathrm{rad}$ and latitude $\beta = +0.98\,\mathrm{rad}$, polarization angle $\psi = 2.70\,\mathrm{rad}$, and inclination $\iota = 0.58\,\mathrm{rad}$ ($\approx 33^\circ$).

\paragraph{Bayesian model.}
We assume a search has localized the source to a small sky patch, so we fix the sky position and orientation at their injected values and sample only $(f_0,\dot f,A,\phi_0)$. The $(f_0,\dot f)$ priors are Gaussian, recentered on the injected search-estimate values, with widths set in units of the resolution elements $1/T$ and $1/T^2$ so the setup is baseline-independent. The $f_0$ prior is narrow, $\sigma_{f_0}=0.32/T\approx1.5\times10^{-8}\,\mathrm{Hz}$, reflecting that the search has already localized $f_0$ to within a frequency bin, while the $\dot f$ prior is deliberately broad, $\sigma_{\dot f}=10/T^2\approx2.1\times10^{-14}\,\mathrm{Hz\,s^{-1}}$, so that $\dot f$ is constrained by the data rather than the prior.

\begin{figure}[!htbp]
  \centering
  \includegraphics[width=\linewidth]{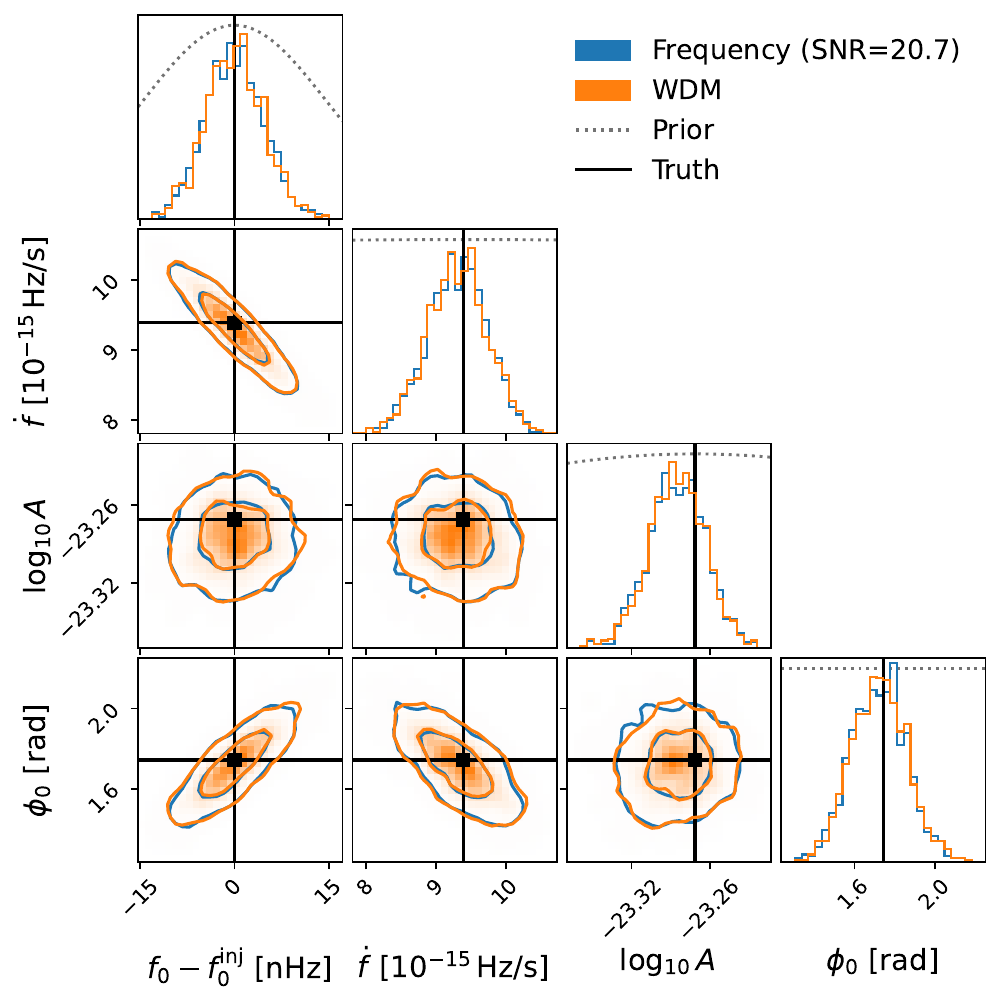}
  \caption{\textbf{Frequency-domain versus WDM-domain posterior.}
    Overlaid corner plot for the two analyses of the same realization, with
    frequency in blue and WDM in orange, the injected truth marked as black solid line, and the
    prior shown as dotted line in the posterior marginals. 
    The two posteriors overlap closely in
    $(f_0,\dot f,A,\phi_0)$, and both are far narrower than the prior.
    }
  \label{fig:lisa_gb_corner}
  \script{lisa_gb_corner.py}
\end{figure}

For numerical conditioning we replace $(A,\phi_0)$ by the Cartesian amplitude coordinates
\begin{equation}
  g_c = A\cos\phi_0, \qquad g_s = A\sin\phi_0 .
  \label{eq:gc_gs}
\end{equation}
At fixed $(f_0,\dot f)$ the waveform is linear in $(g_c,g_s)$, $\tilde h = g_c\tilde h_c + g_s\tilde h_s$, which makes the model linear in the amplitude coordinates and avoids both the strong $A$--$\phi_0$ posterior correlation and the periodic boundary in $\phi_0$. We place an isotropic Gaussian prior on $(g_c,g_s)$, equivalent to a Rayleigh prior on $A$ and an exactly uniform prior on $\phi_0$. All four sampled coordinates are standardized to order unity. These priors are far wider than the posteriors they yield: in Figure~\ref{fig:lisa_gb_corner} the prior is nearly flat across each posterior, most visibly for $\dot f$ whose posterior is tens of times narrower than its prior, confirming that the data drive the inference.

The frequency-domain analysis evaluates the Whittle likelihood in a local A/E/T
Fourier band centered on the source prior support,
\begin{widetext}
\begin{equation}
  \log \mathcal{L}_{\rm freq}(\boldsymbol{\theta}) \propto {} -\frac{2\Delta t}{N}
  \sum_{c\in\{A,E,T\}} \sum_{l = 1}^{N/2 - 1}  \frac{\bigl|\tilde d^c[l]-\tilde h^c[l;\boldsymbol{\theta}]\bigr|^2}
           {S^c[l]},
\end{equation}
where $\tilde d^c[l]$ and $\tilde h^c[l;\boldsymbol{\theta}]$ are the data and
template Fourier modes in channel $c$ and bin $l$, and $S^c[l]$ is the stationary
PSD. The WDM analysis starts from the same Fourier patch, embeds the
template into it, and applies the WDM transform with $N_t=32, N_f = 4096$, $a=1/3$, and
$d=1$, giving
\begin{equation}
  \log \mathcal{L}_{\rm WDM}(\boldsymbol{\theta})
  = -\frac{1}{2}\sum_{c\in\{A,E,T\}}\sum_{\substack{n \\ m\in\mathrm{int}}}
    \frac{\bigl(w^{(d)}_{c,nm}-w^{(h)}_{c,nm}(\boldsymbol{\theta})\bigr)^2}{S^c_{nm}},
\end{equation}
\end{widetext}
with stationary per-pixel variance $S^{c}_{nm} = N\,S_c[mN_t/2]/(2\Delta t)$.
Both analyses thus use the same stationary PSD and differ only in the likelihood
representation. Note that we have dropped the two noise terms in both likelihood functions since we are estimating parameters of the signal model rather than the noise model. 

Both posteriors are sampled with the NumPyro~\cite{phan2019composable} implementation of the No-U-Turn Sampler (NUTS). We use the standardized coordinates and fix the NUTS mass matrix to the inverse Fisher information evaluated at the reference parameters, which supplies a stable metric for the narrow, strongly correlated $(f_0,\dot f)$ posterior ridge. Chains are initialized near the reference with a small Fisher-scaled jitter. We use target acceptance probability $0.9$, maximum tree depth $10$, $1000$ warmup steps, and $1500$ retained samples per chain across two chains. For all reported results we require the Gelman-Rubin statistic to be $\hat R < 1.1$ in every sampled coordinate and zero divergences.

\begin{figure}[!htbp]
  \centering
  \includegraphics[width=\linewidth]{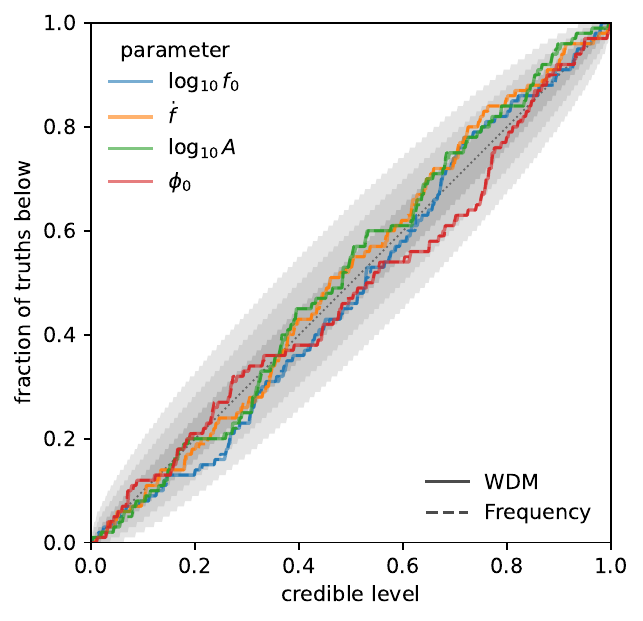}
  \caption{\textbf{Population PP plot.}
    Empirical distribution of the injected truth's posterior quantile across the
    100 seeds, for each parameter, in the WDM domain (solid, lower opacity) and
    frequency domain (dashed). Grey shading shows the nested $1/2/3\sigma$
    pointwise confidence bands expected under the uniform null. The two domains
    track each other closely and every parameter remains within the $3\sigma$
    band. The smallest per-parameter Kolmogorov--Smirnov $p$-value across both
    domains is $\approx0.25$ (for $\phi_0$), consistent with uniformity.}
  \label{fig:lisa_gb_pp}
  \script{lisa_gb_pp.py}
\end{figure}

\paragraph{Results.}
We run the inference for 100 independent seeds, drawing new source parameters, sky position, orientation, target SNR (uniform in $[20,45]$), and noise realization for each. The two posteriors for one representative seed are overlaid in Figure~\ref{fig:lisa_gb_corner}, where the frequency-domain and WDM-domain contours coincide and both contain the injected truth. To quantify this across the population we compare the WDM-domain and frequency-domain one-dimensional marginal posteriors with the Jensen--Shannon divergence (JSD), estimated from the posterior samples with a kernel-density estimator.
Across all 100 seeds and all four sampled parameters the per-seed JSD has a median of $5\times10^{-6}\,\mathrm{bits}$ and a $95$th percentile of $1.4\times10^{-3}\,\mathrm{bits}$, and the posterior-median offsets between the two domains are uniformly small compared with the marginal widths. The population PP curves in Figure~\ref{fig:lisa_gb_pp} likewise track closely between the two domains. These results confirm that, under a shared stationary noise model, the WDM-domain likelihood faithfully reproduces the frequency-domain posterior for resolved-GB inference.

\section{Conclusions}\label{sec:conclusions}

We have presented a self-contained, pedagogical account of the Wilson--Daubechies--Meyer (WDM) wavelet-packet transform, building on the theoretical foundations established in prior work~\cite{Necula_2012,Klimenko:2005xv,Klimenko:2015ypf,Cornish:2020odn}, and its implementation in the open-source \package\ package. The primary goal of this work is to make the discrete, sampled WDM implementation explicit and reproducible for practitioners new to the transform, by walking through the full three-case basis definition (DC, interior, and Nyquist channels), the packed $(N_t, N_f{+}1)$ coefficient layout, the forward and inverse transforms with detailed derivations, and the conditions under which WDM coefficients may be treated as approximately independent.

On the implementation side, the NumPy and JAX backends both achieve exact round-trip reconstruction to floating-point precision. The JAX backend provides substantially lower runtimes for large transforms, making it practical to incorporate WDM-domain operations into inference pipelines that require millions of likelihood evaluations.

Several limitations of the current work should be noted. First, the package implements only the $d=1$ Meyer window; higher-order windows ($d > 1$) offer better spectral leakage suppression and may be preferable for data with sharp spectral features.
Second, the paper does not address the generation of gravitational-wave signals directly in the WDM domain — where much of the computational advantage of the WDM representation lies~\cite{Cornish:2020odn} — and we defer fast WDM-domain waveform generation to future work. Third, the treatment of data gaps and edge effects, while conceptually straightforward, requires careful handling of the boundary basis elements and is left to downstream analyses built on top of the package.%

Fourth, the LISA galactic-binary inference example of Section~\ref{sec:lisa_gb} uses a time-averaged effective PSD in both the frequency-domain and WDM-domain likelihoods; this controlled comparison isolates the WDM basis change but does not exercise the principal advantage of the WDM representation, namely the ability to assign a time-varying WDM domain PSD $S_{nm}$ to each time pixel and thereby accommodate genuinely non-stationary noise without off-diagonal covariance terms.
Demonstrating this advantage on realistic LISA non-stationary backgrounds is an important direction for
future work.

Looking ahead, the most immediate extensions are the incorporation of fast WDM-domain waveform generation for compact binary signals, the implementation of evolutionary power spectral density estimation directly on the WDM grid, and integration of the package into existing LISA data analysis frameworks.  The combination of a pedagogical reference, a tested open-source implementation, and GPU-accelerated performance should lower the barrier to adopting the WDM transform within modern, autodiff-compatible Python inference pipelines for the long-duration, non-stationary signals expected from future detectors such as LISA.
Because the JAX backend is fully differentiable, gradient-based samplers (HMC, NUTS) and variational inference are available out of the box on top of the WDM likelihood, without any additional differentiation work by the user. Future work should focus on determining the optimal tiling of the WDM space for each source class. A systematic comparison with the short-time Fourier transform, assessing respective advantages and limitations, would help identify the most appropriate time–frequency representation and tiling for different types of sources. 

\section*{Data and Software Availability}
The software developed for this research is archived on Zenodo with a
version-specific DOI~\cite{wdm_transform_zenodo}, which should be cited when
using the package.
The source code is also hosted on GitHub at
\url{https://github.com/pywavelet/wdm_transform} and released on PyPI
(installable with \texttt{pip install wdm\_transform}). Documentation and worked
examples are available in the repository documentation at
\url{https://github.com/pywavelet/wdm_transform/tree/main/docs}.
The online documentation includes minimal examples demonstrating the construction of a WDM tiling, application of the forward transform, and round-trip reconstruction with the inverse transform. The in-text code links (\scalebox{0.65}{\faGithub}) are pinned to the tagged release \texttt{v0.5.0} of the package, so that the referenced line numbers remain valid as the codebase evolves.

\begin{acknowledgments}
    The authors thank Matthew Digman, Neil Cornish, Will Farr, Aaron Johnson and Solano Sousa Felicio for discussions.
    We also express our gratitude to Astrid Lamberts for organizing the ``Towards LISA catalogs'' workshop (Nice, 2023), and to Tyson B. Littenberg for organizing the 3rd LISA Sprint meeting (Huntsville, 2025), where this project was further developed.
    O.~Burke sincerely appreciates insightful discussions with Christopher Moore, Christian Chapman-Bird, and Sylvain Marsat during the preparation of this work.
    AV gratefully acknowledges support from the Marsden Fund Council grants MFP-UOA2131 and MFP-UOA2531,
	funded by the New Zealand Government and managed by the Royal Society Te Apārangi.
    GM acknowledges support from the ANR grant AAPG2024 PRC - GalaxyFIT. 
    O.~Burke acknowledges financial support from the Grant UKRI972 awarded via the UK Space Agency.  
    L. Speri acknowledges support through the European Space Agency (ESA) Research Fellowship in Space Science.
    Part of this work was performed on the OzSTAR national facility at Swinburne
	University of Technology. The OzSTAR program receives funding in part from the Astronomy National Collaborative Research Infrastructure Strategy (NCRIS) allocation provided by the Australian Government, and from the Victorian Higher Education State Investment Fund (VHESIF) provided by the Victorian Government.
\end{acknowledgments}

\appendix
\onecolumngrid

\section{Orthonormality condition of the \texorpdfstring{$g_{nm}$}{g_{nm}} basis}\label{app:normality_condition}
To remind the reader, we introduce the basis $\tilde g_{nm}(f)$, given by Eq.~\eqref{eq:gnm_full_discrete},
\begin{align}
\label{eq:gnm_discr}
\tilde{g}_{nm}[l] =
\dfrac{1}{\sqrt{2}}\begin{cases}
e^{-4\pi i n l/N_t}\, \tilde{\varphi}[l], & m=0, \\[6pt]
e^{-2\pi i n l/N_t}\Big(C_{nm}\,\tilde{\varphi}[l - mN_t/2] + C_{nm}^*\,\tilde{\varphi}[l + mN_t/2]\Big), & 0 < m < N_f, \\[6pt]
e^{-4\pi i n l/N_t}\Big(\tilde{\varphi}[l - N/2] + \tilde{\varphi}[l + N/2]\Big), & m = N_f,
\end{cases}
\end{align}
The closed-form expressions below are written
for $d=1$ to keep the algebra explicit. We do remark, however, that the orthonormality and
normalisation results extend to arbitrary $d$ using the incomplete-beta
identity $\nu_d(1-y)+\nu_d(y)=1$ (see Eq.~\eqref{eq:nu_d_def}). We restrict to
$d=1$ here purely for pedagogical clarity\footnote{Analogous proofs of the following identities and formulas can be still obtained in the $d>1$ case by exploiting the symmetry properties of the incomplete beta function}:
\begin{align}\label{eq:phi_discrete}
\tilde\varphi[l]&=\sqrt{\frac{2}{N_t}}\begin{cases}
1 & |l_r|<A \\
\cos \left[\frac{\pi}{2}\,\left(
\frac{|l_r|-A}{B}\right)\right] & A \leq|l_r| < A+B \\
0 & |l_r|\geq A+B
\end{cases}\nonumber\\
l_r&=l\frac{2}{N_t}
\end{align}
where $l\in\{-\frac{N_t}{2},-\frac{N_t}{2}+1,...,\frac{N_t}{2}-1\}$. Now we verify that
\begin{align}
\sum_{nm}&\tilde{g}_{nm}[l]\tilde{g}_{nm}^*[l']=\frac{1}{2}\sum_n e^{-4\pi i n (l-l')/N_t}\, \tilde{\varphi}[l]\,\tilde{\varphi}[l']\nonumber\\
&+\frac{1}{2}\sum_{0<m<N_f}\sum_{n}e^{-2\pi i n (l-l')/N_t}\,\Big(C_{nm}\,\tilde{\varphi}[l - mN_t/2] + C_{nm}^*\,\tilde{\varphi}[l + mN_t/2]\Big)\nonumber\\
&\hspace{10em}\times\Big(C_{nm}^*\,\tilde{\varphi}[l' - mN_t/2] + C_{nm}\,\tilde{\varphi}[l' + mN_t/2]\Big)\nonumber\\
&+\frac{1}{2}\sum_n e^{-4\pi i n (l-l')/N_t}\Big(\tilde{\varphi}[l - N/2] + \tilde{\varphi}[l + N/2]\Big)\Big(\tilde{\varphi}[l' - N/2] + \tilde{\varphi}[l' + N/2]\Big)
\end{align}
Now we simplify the $0<m<N_f$ term, by noticing that $C_{nm}^2=C_{nm}^{*2}=(-1)^{m}(-1)^{n}$ and $|C_{nm}|^2=1$
\begin{align}
\frac{1}{2}\sum_{0<m<N_f}\sum_{n}&e^{-2\pi i n (l-l')/N_t}\,\Big(\tilde{\varphi}[l - mN_t/2]\tilde{\varphi}[l' - mN_t/2] + \tilde{\varphi}[l + mN_t/2]\tilde{\varphi}[l' + mN_t/2]\nonumber\\
&+ (-1)^{m}(-1)^{n}\tilde{\varphi}[l - mN_t/2]\tilde{\varphi}[l' + mN_t/2]+ (-1)^{m}(-1)^{n}\tilde{\varphi}[l + mN_t/2]\tilde{\varphi}[l' - mN_t/2]\Big)
\end{align}
one notices that the summation over $n$ can be directly performed, and it vanishes for every value but $l=l'$. In particular, the second line (the mixed terms) always vanishes. A similar result holds for the $m=0,N_f$ cases, therefore we can write
\begin{align}
\sum_{nm}&\tilde{g}_{nm}[l]\tilde{g}_{nm}^*[l']=\delta_{ll'}\frac{N_t}{2}\left(\tilde\varphi^2[l]+\sum_{0<m<N_f}\tilde\varphi^2[l-mN_t/2]+\sum_{0<m<N_f}\tilde\varphi^2[l+mN_t/2]+\tilde\varphi^2[l-N/2]+\tilde\varphi^2[l+N/2]\right)
\end{align}
Now the thing to notice is that each value of $l$ lies in two and only two of these windows. Let $\bar m$ and $\bar m+1$ be the indices of these two windows, such that $\frac{\bar m N_t}{2}<|l|<\frac{(\bar m+1)N_t}{2}$ ($\bar m$ could be 0, or $\bar m+1$ can be $N_f$). Then the sum becomes
\begin{align}
\sum_{nm}&\tilde{g}_{nm}[l]\tilde{g}_{nm}^*[l']=\delta_{ll'}\frac{N_t}{2}\left(\tilde\varphi^2[l-\bar m N_t/2]+\tilde\varphi^2[l-(\bar m+1)N_t/2]\right)
\end{align}

Let $l\geq 0$ for sake of simplicity ($l<0$ cases are analogous). Then either $l<\frac{(m+A)N_t}{2}$ or $l\geq\frac{(m+A)N_t}{2}$.
In the first case, we have $\tilde\varphi^2[l-\bar m N_t/2]=\frac{2}{N_t}$ and $\tilde\varphi^2[l-(\bar m+1)N_t/2]=0$, since for $|\frac{2l}{N_t}-(\bar m+1)|>A+B$. In the second case, we have
\begin{align}
\sum_{nm}&\tilde{g}_{nm}[l]\tilde{g}_{nm}^*[l']=\delta_{ll'}\left(\cos^2\left[\frac{\pi}{2}\frac{\frac{2l}{N_t}-\bar m -A}{B}\right]+\cos^2\left[\frac{\pi}{2}\frac{\bar m + 1 -\frac{2l}{N_t} -A}{B}\right]\right)
\end{align}
which after some algebra simplifies to
\begin{align}\label{eq:last_step_gnm_orthonorm}
\sum_{nm}&\tilde{g}_{nm}[l]\tilde{g}_{nm}^*[l']=\delta_{ll'}\left(1+\cos\left(\frac{\pi}{2}\frac{1-2A}{B}\right)\cos\left(2\frac{\pi}{2}\frac{\frac{2l}{N_t}-\bar m -A}{B}-\frac{\pi}{2}\frac{1-2A}{B}\right)\right)=\delta_{ll'}
\end{align}
where we used the fact of $2A+B=1$. Therefore we find that for every value of $l$ we have%
\begin{align}
\sum_{nm}&\tilde{g}_{nm}[l]\tilde{g}_{nm}^*[l']=\delta_{ll'}
\end{align}
\subsection{The second (quasi)normality condition}
We want to evaluate
\begin{align}
\sum_{l}&\tilde{g}_{nm}[l]\tilde{g}_{pq}^*[l]
\end{align}
we immediately note that, since the window support in frequency is $N_t/2$ we are forced to have $q=m$. Now we have to consider three cases.

\textbf{Case $m=0$}, for which we have
\begin{align}
\sum_{l}\tilde{g}_{n0}[l]\tilde{g}_{p0}^*[l]&=\frac{1}{2}\sum_{l=-N/2}^{N/2-1} e^{-4\pi i l (n-p)/N_t}\, \tilde{\varphi}^2[l]=\frac{1}{2}\sum_{l=-N_t/2}^{N_t/2-1} e^{-4\pi i l (n-p)/N_t}\, \tilde{\varphi}^2[l]=\nonumber\\
&=\sum_{l=-N_t/2}^{-1}\left(e^{-4\pi i l (n-p)/N_t}\tilde{\varphi}^2[l]+e^{-4\pi i (l+N_t/2) (n-p)/N_t}\tilde{\varphi}^2[l+N_t/2]\right)
\end{align}
but now we note that $e^{-4\pi i (l+N_t/2) (n-p)/N_t}=e^{-4\pi i l(n-p)/N_t}$ and in the previous subsection of this appendix we have also shown that for $|l|\leq N_t/2$ we have $\tilde{\varphi}^2[l]+\tilde{\varphi}^2[l+N_t/2]=\frac{2}{N_t}$, therefore the last expression simplifies to
\begin{align}
\sum_{l}\tilde{g}_{n0}[l]\tilde{g}_{p0}^*[l]&=\frac{1}{N_t}\sum_{l=-N_t/2}^{-1}e^{-4\pi i l (n-p)/N_t}
\end{align}
this is a geometric series in $l$, which vanishes for every choice of $n-p$, but the only two cases $n-p=0$ and $n-p=\pm N_t/2$, for which the exponential evaluates to $1$, therefore obtaining,
\begin{align}
\sum_{l}\tilde{g}_{n0}[l]\tilde{g}_{p0}^*[l]&=\frac{1}{2}\left(\delta_{n,p}+\delta_{n,p\pm \frac{N_t}{2}}\right)
\end{align}

\textbf{Case $m=N_f$}, for which we have
\begin{align}
\sum_{l}\tilde{g}_{n N_f}[l]\tilde{g}_{p N_f}^*[l]&=\frac{1}{2}\sum_{l=-N/2}^{N/2-1} e^{-4\pi i l (n-p)/N_t}\, \left(\tilde{\varphi}[l-N/2]+\tilde{\varphi}[l+N/2]\right)^2=\nonumber\\
&=\frac{1}{2}\sum_{l=-N/2}^{N/2-1} e^{-4\pi i l (n-p)/N_t}\, \left(\tilde{\varphi}^2[l-N/2]+\tilde{\varphi}^2[l+N/2]\right)=\nonumber\\
&=\frac{1}{2}\sum_{l=-N}^{-1} e^{-4\pi i (l+N/2)(n-p)/N_t}\tilde{\varphi}^2[l]+\frac{1}{2}\sum_{l=0}^{N-1} e^{-4\pi i (l-N/2)(n-p)/N_t}\tilde{\varphi}^2[l]=\nonumber\\
&=\frac{1}{2}\sum_{l=-N_t/2}^{-1} e^{-4\pi i l(n-p)/N_t}\tilde{\varphi}^2[l]+\frac{1}{2}\sum_{l=0}^{N_t/2-1} e^{-4\pi i l(n-p)/N_t}\tilde{\varphi}^2[l]=\nonumber\\
&=\frac{1}{2}\sum_{l=-N_t/2}^{N_t/2-1} e^{-4\pi i l(n-p)/N_t}\tilde{\varphi}^2[l]
\end{align}
and from here the proof is identical to the $m=0$ case, leading to
\begin{align}
\sum_{l}\tilde{g}_{nN_f}[l]\tilde{g}_{pN_f}^*[l]&=\frac{1}{2}\left(\delta_{n,p}+\delta_{n,p\pm \frac{N_t}{2}}\right)
\end{align}

\textbf{Case $0<m<N_f$}, for which we have
\begin{align}
\sum_{l}\tilde{g}_{nm}[l]\tilde{g}_{pm}^*[l]&=\frac{1}{2}\sum_{l=-N/2}^{N/2-1} \Bigg[e^{-2\pi i l (n-p)/N_t}\,\Big(C_{nm}\,\tilde{\varphi}[l - mN_t/2] + C_{nm}^*\,\tilde{\varphi}[l + mN_t/2]\Big) \\ 
& \qquad \times \Big(C_{pm}^*\,\tilde{\varphi}[l - mN_t/2] + C_{pm}\,\tilde{\varphi}[l + mN_t/2]\Big) \Bigg]\nonumber\\
&=\frac{1}{2}\sum_{l=-N/2}^{N/2-1} e^{-2\pi i l (n-p)/N_t}\,\Big(C_{nm}C^*_{pm}\,\tilde{\varphi}^2[l - mN_t/2] + C_{nm}^*C_{pm}\,\tilde{\varphi}^2[l + mN_t/2]\Big)
\end{align}
Let's take the first term and shift the frequency, rewriting
\begin{align}
\frac{1}{2}C_{nm}C^*_{pm}\sum_{l=-N_t/2}^{N_t/2-1} e^{-2\pi i (l+mN_t/2) (n-p)/N_t}\tilde{\varphi}^2[l]=\frac{1}{2}C_{nm}C^*_{pm}e^{-\pi i m(n-p)}\sum_{l=-N_t/2}^{N_t/2-1} e^{-2\pi i l(n-p)/N_t}\tilde{\varphi}^2[l]
\end{align}
and also in this case, from here the proof is identical to the $m=0$ case, but where in this case, given the different exponent, the series is nonvanishing for $n=p$ and not for $n=p\pm N_t/2$. The expression then simplifies to
\begin{align}
\frac{1}{2}C_{nm}C^*_{pm}\sum_{l=-N_t/2}^{N_t/2-1} e^{-2\pi i (l+mN_t/2) (n-p)/N_t}\tilde{\varphi}^2[l]=\frac{1}{2}C_{nm}C^*_{nm}=\frac{1}{2}
\end{align}
an analogous proof holds for the term with $+mN_t/2$, therefore the full result is that in this case, for $0<m<N_f$, we find
\begin{align}
\sum_{l}\tilde{g}_{nm}[l]\tilde{g}_{pm}^*[l]=\delta_{n,p}
\end{align}
To wrap up all the cases we find
\begin{align}
\sum_{l}\tilde{g}_{nm}[l]\tilde{g}_{pq}^*[l]=\delta_{m,q}\begin{cases}
    \frac{1}{2}\left(\delta_{n,p}+\delta_{n,p\pm \frac{N_t}{2}}\right) \qquad m=\{0,N_f\}\\
    \delta_{n,p} \qquad 0<m<N_f
\end{cases}
\end{align}

\subsection{Normalization of the Meyer window}\label{app:normalization_Meyer}

We verify that the Meyer window defined in
Eq.~\eqref{eq:phi_discrete} is correctly normalized. We compute
\begin{align}
\sum_{l=-N/2}^{N/2-1} \tilde{\varphi}^2[l]&=2\sum_{l=0}^{A\frac{N_t}{2}-1}\frac{2}{N_t}-\frac{2}{N_t}+2\sum_{l=A\frac{N_t}{2}}^{(A+B)\frac{N_t}{2}-1}\frac{2}{N_t}\cos^2\left[\frac{\pi}{2}\frac{\frac{2l}{N_t}-A}{B}\right]=2A-\frac{2}{N_t}+\frac{4}{N_t}\sum_{k=0}^{B\frac{N_t}{2}-1}\cos^2\left[\frac{\pi}{2}\frac{2k}{BN_t}\right]=\nonumber\\
&=2A-\frac{2}{N_t}+\frac{4}{N_t}\sum_{k=0}^{B\frac{N_t}{2}-1}\frac{1}{2}+\frac{4}{N_t}\sum_{k=0}^{B\frac{N_t}{2}-1}\frac{1}{2}\cos\left(\frac{2\pi k}{BN_t}\right)=2A-\frac{2}{N_t}+B+\frac{2}{N_t}=1
\end{align}
where on the first line the term $-\frac{2}{N_t}$ avoids double counting $l=0$, where the last summation equals $+\frac{2}{N_t}$ after performing the geometrical series and where we remember that $2A+B=1$ is a property of the window function.
Please note that a similar proof can be obtained for Meyer windows with $d>1$ by exploiting the symmetry property of the incomplete beta functions. Also in this case, we omit the general proof for the sake of pedagogical clarity.

\section{Detailed derivation for the forward and inverse transforms}\label{app:forward_derivation}

 This appendix provides the full algebraic steps leading to the
compact expressions for the forward WDM coefficients presented in
Section~\ref{sec:wdm}. We work in the Fourier domain throughout,
writing the forward transform as
\begin{align}
  w_{nm}=\sum_{l=-N/2}^{N/2-1} \tilde{x}[l]\,\tilde{g}_{nm}^*(f[l])\,,
\end{align}
where the discrete frequencies are mapped as $f[l]=l\,\Delta f=l/T$.

\subsection{Interior channels \texorpdfstring{$(0<m<N_f)$}{(0<m<N_f)}}

Inserting the interior-channel basis element from
Eq.~\eqref{eq:gnm_full_discrete} and splitting the sum over positive and
negative frequencies gives
\begin{align}
w_{nm}&=\frac{1}{\sqrt{2}}\sum_{l=0}^{N/2-1} \tilde{x}[l]\, e^{2\pi iln/N_t}\left[C_{nm}^*\tilde{\varphi}\!\left[l-\tfrac{mN_t}{2}\right]+C_{nm}\tilde{\varphi}\!\left[l+\tfrac{mN_t}{2}\right]\right]\nonumber\\
&+\frac{1}{\sqrt{2}}\sum_{l=-N/2}^{-1} \tilde{x}[l]\, e^{2\pi iln/N_t}\left[C_{nm}^*\tilde{\varphi}\!\left[l\!-\!\tfrac{mN_t}{2}\right]+C_{nm}\tilde{\varphi}\!\left[l\!+\!\tfrac{mN_t}{2}\right]\right].
\end{align}
Using the reality
condition $\tilde{x}[-l] = \tilde{x}[l]^*$ together with the
symmetry $\tilde{\varphi}[-l] = \tilde{\varphi}[l]$, the second sum
is the complex conjugate of the first. The boundary terms at $l=0$
and $l=-N/2$ vanish because the window support does not extend to
those frequencies for interior channels. We therefore obtain
\begin{align}\label{eq:app_wnm_general}
  w_{nm}&=\sqrt{2}\,\Re\sum_{l=1}^{N/2-1} \tilde{x}[l]\, e^{2\pi
  iln/N_t}\,C_{nm}^*\,\tilde{\varphi}\!\left[l-\tfrac{mN_t}{2}\right]\nonumber\\
  &=\sqrt{2}\,(-1)^{nm}\,\Re\sum_{l'=-N_t/2}^{N_t/2-1}
  \tilde{x}[l'+mN_t/2]\, e^{2\pi
  il'n/N_t}\,C_{nm}^*\,\tilde{\varphi}[l']\,,
\end{align}
where the last equality follows from the substitution $l' = l -
mN_t/2$ and the fact that $\tilde{\varphi}[l']$ is nonzero only
for $|l'| \leq N_t/2$. This is Eq.~\eqref{eq:wnm_interior} of the
main text, and corresponds (up to normalization conventions) to
Eq.~17 of ~\cite{Cornish:2020odn}.

\subsection{DC channel \texorpdfstring{$(m = 0)$}{(m=0)}}

For the DC edge channel, the basis element has a doubled time-shift
exponent and no $C_{nm}$ factor. The same positive/negative frequency
splitting yields
\begin{align}
  w_{n0}&=\frac{1}{\sqrt{2}}\sum_{l=0}^{N/2-1} \tilde{x}[l]\, e^{4\pi
  iln/N_t}\,\tilde{\varphi}[l]
  +\frac{1}{\sqrt{2}}\sum_{l=-N/2}^{-1} \tilde{x}[l]\, e^{4\pi
  iln/N_t}\,\tilde{\varphi}[l]
  \nonumber\\
  &=\,\sqrt{2}\Re\left[\sum_{l=1}^{N_t/2-1} \tilde{x}[l]\, e^{4\pi
  iln/N_t}\,\tilde{\varphi}[l]\right]+\frac{1}{\sqrt{2}}\tilde{x}[0]\,\tilde{\varphi}[0]\,,
\end{align}
which is Eq.~\eqref{eq:wn0} of the main text.
\subsection{Nyquist channel \texorpdfstring{$(m = N_f)$}{(m=N_f)}}

\begin{align}
w_{nm}&=\frac{1}{\sqrt{2}}\sum_{l=0}^{N/2-1} \tilde{x}[l]\, e^{4\pi iln/N_t}\left[\tilde{\varphi}\!\left[l-\tfrac{N}{2}\right]+\tilde{\varphi}\!\left[l+\tfrac{N}{2}\right]\right]+
\frac{1}{\sqrt{2}}\sum_{l=-N/2}^{-1} \tilde{x}[l]\, e^{4\pi iln/N_t}\left[\tilde{\varphi}\!\left[l\!-\!\tfrac{N}{2}\right]+\tilde{\varphi}\!\left[l\!+\!\tfrac{N}{2}\right]\right]\nonumber\\
&=\frac{1}{\sqrt{2}}\sum_{l=0}^{N/2-1} \tilde{x}[l]\, e^{4\pi iln/N_t}\left[\tilde{\varphi}\!\left[l-\tfrac{N}{2}\right]+\tilde{\varphi}\!\left[l+\tfrac{N}{2}\right]\right]+
\frac{1}{\sqrt{2}}\sum_{l=1}^{N/2} \tilde{x}[-l]\, e^{-4\pi iln/N_t}\left[\tilde{\varphi}\!\left[-l\!-\!\tfrac{N}{2}\right]+\tilde{\varphi}\!\left[-l\!+\!\tfrac{N}{2}\right]\right].
\end{align}
The $l=0$ term in the first summation vanishes, and the $l=N/2$ term is treated separately:
\begin{align}
  w_{nm}&=\sqrt{2}\,\Re\sum_{l=1}^{N/2-1} \tilde{x}[l]\, e^{4\pi
  iln/N_t}\,\,\tilde{\varphi}\!\left[l-\tfrac{N}{2}\right]+\frac{1}{\sqrt{2}}\tilde{x}[N/2]\, e^{-2\pi
  inN_f}\tilde{\varphi}[0]\nonumber\\
  &=\sqrt{2}\,\,\Re\sum_{l'=-N_t/2}^{-1}
  \tilde{x}[l'+N/2]\, e^{4\pi
  il'n/N_t}\,\tilde{\varphi}[l']+\frac{1}{\sqrt{2}}\tilde{x}[N/2]\, \tilde{\varphi}[0]\,,
\end{align}
assuming an even value for $N_f$.

\subsection{The inverse transform}\label{app:inverse_transform}

We write the inverse transform of the $w_{nm}$ coefficients to recover the signal $\tilde x[l]$ for $l\geq 0$ since we know that for $l<0$ the reality condition holds ($\tilde x[-l]=\tilde x^*[l]$):
\begin{align}
\tilde x[l]=\sum_{n=0}^{N_t-1}\sum_{m=0}^{N_f}w_{nm}g_{nm}[l]&=\frac{1}{\sqrt{2}}\sum_{n}\sum_{0<m<N_f}w_{nm}e^{-2\pi iln/N_t}\left[C_{nm}\tilde{\varphi}\!\left[l-\tfrac{mN_t}{2}\right]+C^*_{nm}\tilde{\varphi}\!\left[l+\tfrac{mN_t}{2}\right]\right]\nonumber\\
&+\frac{1}{\sqrt{2}}\sum_{n}w_{n0}e^{-4\pi iln/N_t}\tilde{\varphi}[l]+\frac{1}{\sqrt{2}}\sum_{n}w_{nN_f}e^{-4\pi iln/N_t}\left[\tilde{\varphi}\!\left[l-\tfrac{N}{2}\right]+\tilde{\varphi}\!\left[l+\tfrac{N}{2}\right]\right]=\nonumber\\
&=\begin{cases}
\tilde x_0[l]\qquad\qquad\qquad \qquad l<N_t/2\\
\tilde x_m[l]\qquad\qquad\qquad m N_t/2\leq l<(m+1)N_t/2\\
\tilde x_{N_f}[l]\qquad\qquad\qquad l \geq N/2-N_t/2
\end{cases}
\end{align}
where we note that for each value of $l$, the inverse transform depends on which of the $m$ bands the $l$ value belongs, in particular
\begin{align}
\tilde x_0[l]&=\frac{1}{\sqrt{2}}\sum_{n}w_{n1}e^{-2\pi iln/N_t}C_{n1}\tilde{\varphi}\!\left[l-\tfrac{N_t}{2}\right]+\frac{1}{\sqrt{2}}\sum_{n}w_{n0}e^{-4\pi iln/N_t}\tilde{\varphi}[l]\nonumber\\
\tilde x_m[l]&=\frac{1}{\sqrt{2}}\sum_n w_{nm}e^{-2\pi iln/N_t}C_{nm}\tilde{\varphi}\!\left[l-\tfrac{mN_t}{2}\right]+\frac{1}{\sqrt{2}}\sum_n w_{n(m+1)}e^{-2\pi iln/N_t}C_{n(m+1)}\tilde{\varphi}\!\left[l-\tfrac{(m+1)N_t}{2}\right]\nonumber\\
\tilde x_{N_f}[l]&=\frac{1}{\sqrt{2}}\sum_n w_{n(N_f-1)}e^{-2\pi iln/N_t}C_{n(N_f-1)}\tilde{\varphi}\!\left[l-\tfrac{N}{2}+\tfrac{N_t}{2}\right]+\frac{1}{\sqrt{2}}\sum_n w_{n N_f}e^{-4\pi iln/N_t}\tilde{\varphi}\!\left[l-\tfrac{N}{2}\right]
\end{align}
Which can be written in a slightly more convenient form:

\begin{align}
\tilde{x}_m[l] = \dfrac{1}{\sqrt{2}}
\begin{cases}
\tilde{\varphi}[l]\, \displaystyle\sum_{n=0}^{N_t - 1} w_{n0}\, e^{-4\pi i n l / N_t},
& m = 0,\ \ 0 \le l < N_t/2, \\[10pt]
\tilde{\varphi}[l - mN_t/2]\, \displaystyle\sum_{n=0}^{N_t - 1} C_{nm}\, w_{nm}\, e^{-2\pi i n l / N_t},
& 1 \le m \le N_f - 1,\ \ (m{-}1)\tfrac{N_t}{2} \le l < (m{+}1)\tfrac{N_t}{2}, \\[10pt]
\tilde{\varphi}[l - N/2]\, \displaystyle\sum_{n=0}^{N_t - 1} w_{nN_f}\, e^{-4\pi i n l / N_t},
& m = N_f,\ \ N/2 - N_t/2 \le l \le N/2,
\end{cases}
\end{align}
which is presented in the main text \eqref{eq:inverse_transform_discrete}. 

\section{The WDM Covariance Matrix in the stationary case}\label{app:stats}

We start from the WDM transform in Eq. \eqref{wnm_def}. It can be re-written in the time domain by injecting the expression of the DFT of $x$ and $g$, which yields
\begin{equation}
\label{eq:wdm-transform-time-domain}
    w_{nm} = N \sum_{k=0}^{N-1} x[k] g_{nm}[k].
\end{equation}
The WDM covariance matrix elements are then
\begin{align}
(\boldsymbol{\Sigma}_w)_{(n,m),(p,q)} &= \mathbb{E}[w_{nm}w_{pq}] = N^2 \sum_{k,k'}g_{nm}[k]g_{pq}[k']\mathbb{E}[x[k]x[k']].
\end{align}
For a wide-sense stationary process, the covariance term is given by the auto-covariance function
\begin{equation}
\label{eq:autocovariance}
R(\kappa) = \mathbb{E}[x[k]x[k+\kappa]],
\end{equation}
which only depends on the time separation between the two data points.
The Wiener-Khinchin theorem states that the Fourier transform of the
time-domain auto-covariance function is related to one-sided power spectral density as
\begin{align}
\label{eq:wiener-khinchin}
    R(\kappa) &= \frac{1}{2}\int_{-\frac{f_s}{2}}^{\frac{f_s}{2}} S(f) e^{2 i \pi f \kappa \Delta t } df
\end{align}
Using Eqs. \eqref{eq:autocovariance} and \eqref{eq:wiener-khinchin} into \eqref{eq:wdm-transform-time-domain} gives
\begin{align}
(\boldsymbol{\Sigma}_w)_{(n,m),(p,q)} &=  \frac{N^2}{2}\int_{-\frac{f_s}{2}}^{\frac{f_s}{2}} S(f)  \sum_{k,k'}g_{nm}[k]g_{pq}[k'] e^{2 i \pi f (k'-k) \Delta t } df \\
& = \frac{N^2}{2}\int_{-\frac{f_s}{2}}^{\frac{f_s}{2}} S(f) \tilde{g}_{nm}(f) \tilde{g}_{pq}^{*}(f) df.
\end{align}
or in the discrete domain
\begin{align}
(\boldsymbol{\Sigma}_w)_{(n,m),(p,q)} \approx \frac{N^2\Delta f}{2}\sum_{l=-N/2}^{N/2 - 1} S[l]\tilde{g}_{nm}[l] \tilde{g}_{pq}^{*}[l]=\frac{N}{2\Delta t}\sum_{l=-N/2}^{N/2 - 1} S[l]\tilde{g}_{nm}[l] \tilde{g}_{pq}^{*}[l]
\end{align}
since $\Delta f=\frac{1}{T}=\frac{1}{N\Delta t}$.
The atoms from the Wilson basis $\tilde{g}_{nm}$ are peaked around the frequency $f_m = m N_t \Delta f/2$. This is where the support of the Wilson basis lies. This means that the dominant contribution of the integral are for those frequencies in the vicinity of $f_m$ and $f_q$. Let us now assume that the PSD varies slowly enough so that we can assimilate it to a constant equal to its value at $(f_m + f_q) / 2$. In that case, the 0th order approximation of the equation above is done by approximating $S\left[l\right]\simeq S\left[(m+q) N_t/ 4\right]$, which now does not depend on $l$ and can be taken out of the summation
\begin{align}
\label{eq:wdm-covariance-stationary}
(\boldsymbol{\Sigma}_w)_{(n,m),(p,q)} & \approx \frac{N}{2\Delta t} S\left[(m+q) N_t/ 4\right]\sum_{l=-N/2}^{N/2 - 1}\tilde{g}_{nm}[l] \tilde{g}_{pq}^{*}[l]=\frac{N}{2\Delta t} S\left[m N_t/ 2\right]\delta_{np}\delta_{mq}.
\end{align}
where the last equality follows from Eq.~\eqref{eq:orthonormality}, in the general case $0<m<N_f$.

\bibliographystyle{apsrev4-2}
\bibliography{bib}

%apsrev4-2.bst 2019-01-14 (MD) hand-edited version of apsrev4-1.bst
%Control: key (0)
%Control: author (72) initials jnrlst
%Control: editor formatted (1) identically to author
%Control: production of article title (-1) disabled
%Control: page (0) single
%Control: year (1) truncated
%Control: production of eprint (0) enabled
\begin{thebibliography}{41}%
\makeatletter
\providecommand \@ifxundefined [1]{%
 \@ifx{#1\undefined}
}%
\providecommand \@ifnum [1]{%
 \ifnum #1\expandafter \@firstoftwo
 \else \expandafter \@secondoftwo
 \fi
}%
\providecommand \@ifx [1]{%
 \ifx #1\expandafter \@firstoftwo
 \else \expandafter \@secondoftwo
 \fi
}%
\providecommand \natexlab [1]{#1}%
\providecommand \enquote  [1]{``#1''}%
\providecommand \bibnamefont  [1]{#1}%
\providecommand \bibfnamefont [1]{#1}%
\providecommand \citenamefont [1]{#1}%
\providecommand \href@noop [0]{\@secondoftwo}%
\providecommand \href [0]{\begingroup \@sanitize@url \@href}%
\providecommand \@href[1]{\@@startlink{#1}\@@href}%
\providecommand \@@href[1]{\endgroup#1\@@endlink}%
\providecommand \@sanitize@url [0]{\catcode `\\12\catcode `\$12\catcode
  `\&12\catcode `\#12\catcode `\^12\catcode `\_12\catcode `\%12\relax}%
\providecommand \@@startlink[1]{}%
\providecommand \@@endlink[0]{}%
\providecommand \url  [0]{\begingroup\@sanitize@url \@url }%
\providecommand \@url [1]{\endgroup\@href {#1}{\urlprefix }}%
\providecommand \urlprefix  [0]{URL }%
\providecommand \Eprint [0]{\href }%
\providecommand \doibase [0]{https://doi.org/}%
\providecommand \selectlanguage [0]{\@gobble}%
\providecommand \bibinfo  [0]{\@secondoftwo}%
\providecommand \bibfield  [0]{\@secondoftwo}%
\providecommand \translation [1]{[#1]}%
\providecommand \BibitemOpen [0]{}%
\providecommand \bibitemStop [0]{}%
\providecommand \bibitemNoStop [0]{.\EOS\space}%
\providecommand \EOS [0]{\spacefactor3000\relax}%
\providecommand \BibitemShut  [1]{\csname bibitem#1\endcsname}%
\let\auto@bib@innerbib\@empty
%</preamble>
\bibitem [{\citenamefont {Królak}\ and\ \citenamefont
  {Trzaskoma}(1996)}]{krolak_1996}%
  \BibitemOpen
  \bibfield  {author} {\bibinfo {author} {\bibfnamefont {A.}~\bibnamefont
  {Królak}}\ and\ \bibinfo {author} {\bibfnamefont {P.}~\bibnamefont
  {Trzaskoma}},\ }\href {https://doi.org/10.1088/0264-9381/13/5/006} {\bibfield
   {journal} {\bibinfo  {journal} {Classical and Quantum Gravity}\ }\textbf
  {\bibinfo {volume} {13}},\ \bibinfo {pages} {813} (\bibinfo {year}
  {1996})}\BibitemShut {NoStop}%
\bibitem [{\citenamefont {Virtuoso}\ and\ \citenamefont
  {Milotti}(2024)}]{Virtuoso2024}%
  \BibitemOpen
  \bibfield  {author} {\bibinfo {author} {\bibfnamefont {A.}~\bibnamefont
  {Virtuoso}}\ and\ \bibinfo {author} {\bibfnamefont {E.}~\bibnamefont
  {Milotti}},\ }\href {https://doi.org/10.1103/PhysRevD.109.102010} {\bibfield
  {journal} {\bibinfo  {journal} {Phys. Rev. D}\ }\textbf {\bibinfo {volume}
  {109}},\ \bibinfo {pages} {102010} (\bibinfo {year} {2024})}\BibitemShut
  {NoStop}%
\bibitem [{\citenamefont {Robinet}\ \emph {et~al.}(2020)\citenamefont
  {Robinet}, \citenamefont {Arnaud}, \citenamefont {Leroy}, \citenamefont
  {Lundgren}, \citenamefont {Macleod},\ and\ \citenamefont
  {McIver}}]{robinet2020}%
  \BibitemOpen
  \bibfield  {author} {\bibinfo {author} {\bibfnamefont {F.}~\bibnamefont
  {Robinet}}, \bibinfo {author} {\bibfnamefont {N.}~\bibnamefont {Arnaud}},
  \bibinfo {author} {\bibfnamefont {N.}~\bibnamefont {Leroy}}, \bibinfo
  {author} {\bibfnamefont {A.}~\bibnamefont {Lundgren}}, \bibinfo {author}
  {\bibfnamefont {D.}~\bibnamefont {Macleod}},\ and\ \bibinfo {author}
  {\bibfnamefont {J.}~\bibnamefont {McIver}},\ }\href
  {https://doi.org/https://doi.org/10.1016/j.softx.2020.100620} {\bibfield
  {journal} {\bibinfo  {journal} {SoftwareX}\ }\textbf {\bibinfo {volume}
  {12}},\ \bibinfo {pages} {100620} (\bibinfo {year} {2020})}\BibitemShut
  {NoStop}%
\bibitem [{\citenamefont {Cornish}\ and\ \citenamefont
  {Littenberg}(2015)}]{Cornish_2015}%
  \BibitemOpen
  \bibfield  {author} {\bibinfo {author} {\bibfnamefont {N.~J.}\ \bibnamefont
  {Cornish}}\ and\ \bibinfo {author} {\bibfnamefont {T.~B.}\ \bibnamefont
  {Littenberg}},\ }\href {https://doi.org/10.1088/0264-9381/32/13/135012}
  {\bibfield  {journal} {\bibinfo  {journal} {Classical and Quantum Gravity}\
  }\textbf {\bibinfo {volume} {32}},\ \bibinfo {pages} {135012} (\bibinfo
  {year} {2015})}\BibitemShut {NoStop}%
\bibitem [{\citenamefont {Chatterji}\ \emph {et~al.}(2004)\citenamefont
  {Chatterji}, \citenamefont {Blackburn}, \citenamefont {Martin},\ and\
  \citenamefont {Katsavounidis}}]{Chatterji:2004qg}%
  \BibitemOpen
  \bibfield  {author} {\bibinfo {author} {\bibfnamefont {S.}~\bibnamefont
  {Chatterji}}, \bibinfo {author} {\bibfnamefont {L.}~\bibnamefont
  {Blackburn}}, \bibinfo {author} {\bibfnamefont {G.}~\bibnamefont {Martin}},\
  and\ \bibinfo {author} {\bibfnamefont {E.}~\bibnamefont {Katsavounidis}},\
  }\href {https://doi.org/10.1088/0264-9381/21/20/024} {\bibfield  {journal}
  {\bibinfo  {journal} {Class. Quant. Grav.}\ }\textbf {\bibinfo {volume}
  {21}},\ \bibinfo {pages} {S1809} (\bibinfo {year} {2004})},\ \Eprint
  {https://arxiv.org/abs/gr-qc/0412119} {arXiv:gr-qc/0412119} \BibitemShut
  {NoStop}%
\bibitem [{\citenamefont {Chassande-Mottin}\ and\ \citenamefont
  {Pai}(2006)}]{ChassandeMottin:2005bm}%
  \BibitemOpen
  \bibfield  {author} {\bibinfo {author} {\bibfnamefont {E.}~\bibnamefont
  {Chassande-Mottin}}\ and\ \bibinfo {author} {\bibfnamefont {A.}~\bibnamefont
  {Pai}},\ }\href {https://doi.org/10.1103/PhysRevD.73.042003} {\bibfield
  {journal} {\bibinfo  {journal} {Phys. Rev. D}\ }\textbf {\bibinfo {volume}
  {73}},\ \bibinfo {pages} {042003} (\bibinfo {year} {2006})},\ \Eprint
  {https://arxiv.org/abs/gr-qc/0512137} {arXiv:gr-qc/0512137} \BibitemShut
  {NoStop}%
\bibitem [{\citenamefont {Gair}\ and\ \citenamefont {Wen}(2005)}]{Gair_2005}%
  \BibitemOpen
  \bibfield  {author} {\bibinfo {author} {\bibfnamefont {J.}~\bibnamefont
  {Gair}}\ and\ \bibinfo {author} {\bibfnamefont {L.}~\bibnamefont {Wen}},\
  }\href {https://doi.org/10.1088/0264-9381/22/18/s49} {\bibfield  {journal}
  {\bibinfo  {journal} {Classical and Quantum Gravity}\ }\textbf {\bibinfo
  {volume} {22}},\ \bibinfo {pages} {S1359–S1371} (\bibinfo {year}
  {2005})}\BibitemShut {NoStop}%
\bibitem [{\citenamefont {Gair}\ \emph {et~al.}(2008)\citenamefont {Gair},
  \citenamefont {Mandel},\ and\ \citenamefont {Wen}}]{Gair:2008ec}%
  \BibitemOpen
  \bibfield  {author} {\bibinfo {author} {\bibfnamefont {J.~R.}\ \bibnamefont
  {Gair}}, \bibinfo {author} {\bibfnamefont {I.}~\bibnamefont {Mandel}},\ and\
  \bibinfo {author} {\bibfnamefont {L.}~\bibnamefont {Wen}},\ }\href
  {https://doi.org/10.1088/0264-9381/25/18/184031} {\bibfield  {journal}
  {\bibinfo  {journal} {Class. Quant. Grav.}\ }\textbf {\bibinfo {volume}
  {25}},\ \bibinfo {pages} {184031} (\bibinfo {year} {2008})},\ \Eprint
  {https://arxiv.org/abs/0804.1084} {arXiv:0804.1084 [gr-qc]} \BibitemShut
  {NoStop}%
\bibitem [{\citenamefont {Speri}\ \emph {et~al.}(2026)\citenamefont {Speri},
  \citenamefont {Tenorio}, \citenamefont {Chapman-Bird},\ and\ \citenamefont
  {Gerosa}}]{Speri_2026}%
  \BibitemOpen
  \bibfield  {author} {\bibinfo {author} {\bibfnamefont {L.}~\bibnamefont
  {Speri}}, \bibinfo {author} {\bibfnamefont {R.}~\bibnamefont {Tenorio}},
  \bibinfo {author} {\bibfnamefont {C.}~\bibnamefont {Chapman-Bird}},\ and\
  \bibinfo {author} {\bibfnamefont {D.}~\bibnamefont {Gerosa}},\ }\bibfield
  {journal} {\bibinfo  {journal} {Physical Review D}\ }\textbf {\bibinfo
  {volume} {113}},\ \href {https://doi.org/10.1103/dh3j-ksfl}
  {10.1103/dh3j-ksfl} (\bibinfo {year} {2026})\BibitemShut {NoStop}%
\bibitem [{\citenamefont {Bandopadhyay}\ \emph {et~al.}(2026)\citenamefont
  {Bandopadhyay}, \citenamefont {Chapman-Bird},\ and\ \citenamefont
  {Vecchio}}]{bandopadhyay2026globaltimefrequencysearchstellarmass}%
  \BibitemOpen
  \bibfield  {author} {\bibinfo {author} {\bibfnamefont {D.}~\bibnamefont
  {Bandopadhyay}}, \bibinfo {author} {\bibfnamefont {C.~E.~A.}\ \bibnamefont
  {Chapman-Bird}},\ and\ \bibinfo {author} {\bibfnamefont {A.}~\bibnamefont
  {Vecchio}},\ }\href {https://arxiv.org/abs/2510.19047} {\bibinfo {title}
  {Global time-frequency search for stellar-mass binary black holes in {LISA}}}
  (\bibinfo {year} {2026}),\ \Eprint {https://arxiv.org/abs/2510.19047}
  {arXiv:2510.19047 [gr-qc]} \BibitemShut {NoStop}%
\bibitem [{\citenamefont {{Colpi}}\ \emph {et~al.}(2024)\citenamefont {{Colpi}}
  \emph {et~al.}}]{2024arXiv240207571C}%
  \BibitemOpen
  \bibfield  {author} {\bibinfo {author} {\bibfnamefont {M.}~\bibnamefont
  {{Colpi}}} \emph {et~al.},\ }\href
  {https://doi.org/10.48550/arXiv.2402.07571} {\bibfield  {journal} {\bibinfo
  {journal} {arXiv e-prints}\ ,\ \bibinfo {eid} {arXiv:2402.07571}} (\bibinfo
  {year} {2024})},\ \Eprint {https://arxiv.org/abs/2402.07571}
  {arXiv:2402.07571 [astro-ph.CO]} \BibitemShut {NoStop}%
\bibitem [{\citenamefont {Luo}\ \emph {et~al.}(2016)\citenamefont {Luo} \emph
  {et~al.}}]{Luo:2015uga}%
  \BibitemOpen
  \bibfield  {author} {\bibinfo {author} {\bibfnamefont {J.}~\bibnamefont
  {Luo}} \emph {et~al.},\ }\href
  {https://doi.org/10.1088/0264-9381/33/3/035010} {\bibfield  {journal}
  {\bibinfo  {journal} {Class. Quant. Grav.}\ }\textbf {\bibinfo {volume}
  {33}},\ \bibinfo {pages} {035010} (\bibinfo {year} {2016})},\ \Eprint
  {https://arxiv.org/abs/1512.02076} {arXiv:1512.02076 [astro-ph.IM]}
  \BibitemShut {NoStop}%
\bibitem [{\citenamefont {Cornish}\ \emph {et~al.}(2021)\citenamefont
  {Cornish}, \citenamefont {Littenberg}, \citenamefont {B\'ecsy}, \citenamefont
  {Chatziioannou}, \citenamefont {Clark}, \citenamefont {Ghonge},\ and\
  \citenamefont {Millhouse}}]{Cornish:2020kdz}%
  \BibitemOpen
  \bibfield  {author} {\bibinfo {author} {\bibfnamefont {N.~J.}\ \bibnamefont
  {Cornish}}, \bibinfo {author} {\bibfnamefont {T.~B.}\ \bibnamefont
  {Littenberg}}, \bibinfo {author} {\bibfnamefont {B.}~\bibnamefont {B\'ecsy}},
  \bibinfo {author} {\bibfnamefont {K.}~\bibnamefont {Chatziioannou}}, \bibinfo
  {author} {\bibfnamefont {J.~A.}\ \bibnamefont {Clark}}, \bibinfo {author}
  {\bibfnamefont {S.}~\bibnamefont {Ghonge}},\ and\ \bibinfo {author}
  {\bibfnamefont {M.}~\bibnamefont {Millhouse}},\ }\href
  {https://doi.org/10.1103/PhysRevD.103.044006} {\bibfield  {journal} {\bibinfo
   {journal} {Phys. Rev. D}\ }\textbf {\bibinfo {volume} {103}},\ \bibinfo
  {pages} {044006} (\bibinfo {year} {2021})},\ \Eprint
  {https://arxiv.org/abs/2011.09494} {arXiv:2011.09494 [gr-qc]} \BibitemShut
  {NoStop}%
\bibitem [{\citenamefont {Necula}\ \emph {et~al.}(2012)\citenamefont {Necula},
  \citenamefont {Klimenko},\ and\ \citenamefont {Mitselmakher}}]{Necula_2012}%
  \BibitemOpen
  \bibfield  {author} {\bibinfo {author} {\bibfnamefont {V.}~\bibnamefont
  {Necula}}, \bibinfo {author} {\bibfnamefont {S.}~\bibnamefont {Klimenko}},\
  and\ \bibinfo {author} {\bibfnamefont {G.}~\bibnamefont {Mitselmakher}},\
  }\href {https://doi.org/10.1088/1742-6596/363/1/012032} {\bibfield  {journal}
  {\bibinfo  {journal} {Journal of Physics: Conference Series}\ }\textbf
  {\bibinfo {volume} {363}},\ \bibinfo {pages} {012032} (\bibinfo {year}
  {2012})}\BibitemShut {NoStop}%
\bibitem [{\citenamefont {{Szczepa{\'n}czyk}}\ \emph
  {et~al.}(2023)\citenamefont {{Szczepa{\'n}czyk}}, \citenamefont {{Salemi}},
  \citenamefont {{Bini}}, \citenamefont {{Mishra}}, \citenamefont {{Vedovato}},
  \citenamefont {{Gayathri}}, \citenamefont {{Bartos}}, \citenamefont
  {{Bhaumik}}, \citenamefont {{Drago}}, \citenamefont {{Halim}}, \citenamefont
  {{Lazzaro}}, \citenamefont {{Miani}}, \citenamefont {{Milotti}},
  \citenamefont {{Prodi}}, \citenamefont {{Tiwari}},\ and\ \citenamefont
  {{Klimenko}}}]{2023PhRvD.107f2002S}%
  \BibitemOpen
  \bibfield  {author} {\bibinfo {author} {\bibfnamefont {M.~J.}\ \bibnamefont
  {{Szczepa{\'n}czyk}}}, \bibinfo {author} {\bibfnamefont {F.}~\bibnamefont
  {{Salemi}}}, \bibinfo {author} {\bibfnamefont {S.}~\bibnamefont {{Bini}}},
  \bibinfo {author} {\bibfnamefont {T.}~\bibnamefont {{Mishra}}}, \bibinfo
  {author} {\bibfnamefont {G.}~\bibnamefont {{Vedovato}}}, \bibinfo {author}
  {\bibfnamefont {V.}~\bibnamefont {{Gayathri}}}, \bibinfo {author}
  {\bibfnamefont {I.}~\bibnamefont {{Bartos}}}, \bibinfo {author}
  {\bibfnamefont {S.}~\bibnamefont {{Bhaumik}}}, \bibinfo {author}
  {\bibfnamefont {M.}~\bibnamefont {{Drago}}}, \bibinfo {author} {\bibfnamefont
  {O.}~\bibnamefont {{Halim}}}, \bibinfo {author} {\bibfnamefont
  {C.}~\bibnamefont {{Lazzaro}}}, \bibinfo {author} {\bibfnamefont
  {A.}~\bibnamefont {{Miani}}}, \bibinfo {author} {\bibfnamefont
  {E.}~\bibnamefont {{Milotti}}}, \bibinfo {author} {\bibfnamefont {G.~A.}\
  \bibnamefont {{Prodi}}}, \bibinfo {author} {\bibfnamefont {S.}~\bibnamefont
  {{Tiwari}}},\ and\ \bibinfo {author} {\bibfnamefont {S.}~\bibnamefont
  {{Klimenko}}},\ }\href {https://doi.org/10.1103/PhysRevD.107.062002}
  {\bibfield  {journal} {\bibinfo  {journal} {\prd}\ }\textbf {\bibinfo
  {volume} {107}},\ \bibinfo {eid} {062002} (\bibinfo {year} {2023})},\ \Eprint
  {https://arxiv.org/abs/2210.01754} {arXiv:2210.01754 [gr-qc]} \BibitemShut
  {NoStop}%
\bibitem [{\citenamefont {{Drago}}\ \emph {et~al.}(2021)\citenamefont
  {{Drago}}, \citenamefont {{Klimenko}}, \citenamefont {{Lazzaro}},
  \citenamefont {{Milotti}}, \citenamefont {{Mitselmakher}}, \citenamefont
  {{Necula}}, \citenamefont {{O'Brian}}, \citenamefont {{Prodi}}, \citenamefont
  {{Salemi}}, \citenamefont {{Szczepanczyk}}, \citenamefont {{Tiwari}},
  \citenamefont {{Tiwari}}, \citenamefont {{Gayathri}}, \citenamefont
  {{Vedovato}},\ and\ \citenamefont {{Yakushin}}}]{2021SoftX..1400678D}%
  \BibitemOpen
  \bibfield  {author} {\bibinfo {author} {\bibfnamefont {M.}~\bibnamefont
  {{Drago}}}, \bibinfo {author} {\bibfnamefont {S.}~\bibnamefont {{Klimenko}}},
  \bibinfo {author} {\bibfnamefont {C.}~\bibnamefont {{Lazzaro}}}, \bibinfo
  {author} {\bibfnamefont {E.}~\bibnamefont {{Milotti}}}, \bibinfo {author}
  {\bibfnamefont {G.}~\bibnamefont {{Mitselmakher}}}, \bibinfo {author}
  {\bibfnamefont {V.}~\bibnamefont {{Necula}}}, \bibinfo {author}
  {\bibfnamefont {B.}~\bibnamefont {{O'Brian}}}, \bibinfo {author}
  {\bibfnamefont {G.~A.}\ \bibnamefont {{Prodi}}}, \bibinfo {author}
  {\bibfnamefont {F.}~\bibnamefont {{Salemi}}}, \bibinfo {author}
  {\bibfnamefont {M.}~\bibnamefont {{Szczepanczyk}}}, \bibinfo {author}
  {\bibfnamefont {S.}~\bibnamefont {{Tiwari}}}, \bibinfo {author}
  {\bibfnamefont {V.}~\bibnamefont {{Tiwari}}}, \bibinfo {author}
  {\bibfnamefont {V.}~\bibnamefont {{Gayathri}}}, \bibinfo {author}
  {\bibfnamefont {G.}~\bibnamefont {{Vedovato}}},\ and\ \bibinfo {author}
  {\bibfnamefont {I.}~\bibnamefont {{Yakushin}}},\ }\href
  {https://doi.org/10.1016/j.softx.2021.100678} {\bibfield  {journal} {\bibinfo
   {journal} {SoftwareX}\ }\textbf {\bibinfo {volume} {14}},\ \bibinfo {eid}
  {100678} (\bibinfo {year} {2021})},\ \Eprint
  {https://arxiv.org/abs/2006.12604} {arXiv:2006.12604 [gr-qc]} \BibitemShut
  {NoStop}%
\bibitem [{\citenamefont {Anderson}\ \emph {et~al.}(2001)\citenamefont
  {Anderson}, \citenamefont {Brady}, \citenamefont {Creighton},\ and\
  \citenamefont {Flanagan}}]{Anderson:2000yy}%
  \BibitemOpen
  \bibfield  {author} {\bibinfo {author} {\bibfnamefont {W.~G.}\ \bibnamefont
  {Anderson}}, \bibinfo {author} {\bibfnamefont {P.~R.}\ \bibnamefont {Brady}},
  \bibinfo {author} {\bibfnamefont {J.~D.~E.}\ \bibnamefont {Creighton}},\ and\
  \bibinfo {author} {\bibfnamefont {E.~E.}\ \bibnamefont {Flanagan}},\ }\href
  {https://doi.org/10.1103/PhysRevD.63.042003} {\bibfield  {journal} {\bibinfo
  {journal} {Phys. Rev. D}\ }\textbf {\bibinfo {volume} {63}},\ \bibinfo
  {pages} {042003} (\bibinfo {year} {2001})},\ \Eprint
  {https://arxiv.org/abs/gr-qc/0008066} {arXiv:gr-qc/0008066} \BibitemShut
  {NoStop}%
\bibitem [{\citenamefont {Klimenko}\ and\ \citenamefont
  {Mitselmakher}(2004)}]{Klimenko:2004qh}%
  \BibitemOpen
  \bibfield  {author} {\bibinfo {author} {\bibfnamefont {S.}~\bibnamefont
  {Klimenko}}\ and\ \bibinfo {author} {\bibfnamefont {G.}~\bibnamefont
  {Mitselmakher}},\ }\href {https://doi.org/10.1088/0264-9381/21/20/025}
  {\bibfield  {journal} {\bibinfo  {journal} {Class. Quant. Grav.}\ }\textbf
  {\bibinfo {volume} {21}},\ \bibinfo {pages} {S1819} (\bibinfo {year}
  {2004})}\BibitemShut {NoStop}%
\bibitem [{\citenamefont {{Digman}}\ and\ \citenamefont
  {{Cornish}}(2023{\natexlab{a}})}]{2023ascl.soft07037D}%
  \BibitemOpen
  \bibfield  {author} {\bibinfo {author} {\bibfnamefont {M.~C.}\ \bibnamefont
  {{Digman}}}\ and\ \bibinfo {author} {\bibfnamefont {N.~J.}\ \bibnamefont
  {{Cornish}}},\ }\href@noop {} {\bibinfo {title} {{WDMWaveletTransforms: Fast
  forward and inverse WDM wavelet transforms}}},\ \bibinfo {howpublished}
  {Astrophysics Source Code Library, record ascl:2307.037} (\bibinfo {year}
  {2023}{\natexlab{a}}),\ \Eprint {https://arxiv.org/abs/2307.037}
  {ascl:2307.037} \BibitemShut {NoStop}%
\bibitem [{\citenamefont {{Digman}}\ and\ \citenamefont
  {{Cornish}}(2023{\natexlab{b}})}]{digman_cornish_bbh_wdm}%
  \BibitemOpen
  \bibfield  {author} {\bibinfo {author} {\bibfnamefont {M.~C.}\ \bibnamefont
  {{Digman}}}\ and\ \bibinfo {author} {\bibfnamefont {N.~J.}\ \bibnamefont
  {{Cornish}}},\ }\href {https://doi.org/10.1103/PhysRevD.108.023022}
  {\bibfield  {journal} {\bibinfo  {journal} {\prd}\ }\textbf {\bibinfo
  {volume} {108}},\ \bibinfo {eid} {023022} (\bibinfo {year}
  {2023}{\natexlab{b}})},\ \Eprint {https://arxiv.org/abs/2212.04600}
  {arXiv:2212.04600 [gr-qc]} \BibitemShut {NoStop}%
\bibitem [{\citenamefont {{Digman}}\ and\ \citenamefont
  {{Cornish}}(2022)}]{digman_cornish_tv_galactic_wdm}%
  \BibitemOpen
  \bibfield  {author} {\bibinfo {author} {\bibfnamefont {M.~C.}\ \bibnamefont
  {{Digman}}}\ and\ \bibinfo {author} {\bibfnamefont {N.~J.}\ \bibnamefont
  {{Cornish}}},\ }\href {https://doi.org/10.3847/1538-4357/ac9139} {\bibfield
  {journal} {\bibinfo  {journal} {\apj}\ }\textbf {\bibinfo {volume} {940}},\
  \bibinfo {eid} {10} (\bibinfo {year} {2022})},\ \Eprint
  {https://arxiv.org/abs/2206.14813} {arXiv:2206.14813 [astro-ph.IM]}
  \BibitemShut {NoStop}%
\bibitem [{\citenamefont {{Pearson}}\ and\ \citenamefont
  {{Cornish}}(2026)}]{pearson_cornish_wdm_gaps}%
  \BibitemOpen
  \bibfield  {author} {\bibinfo {author} {\bibfnamefont {N.}~\bibnamefont
  {{Pearson}}}\ and\ \bibinfo {author} {\bibfnamefont {N.~J.}\ \bibnamefont
  {{Cornish}}},\ }\href {https://doi.org/10.1103/htg6-w4yy} {\bibfield
  {journal} {\bibinfo  {journal} {\prd}\ }\textbf {\bibinfo {volume} {113}},\
  \bibinfo {eid} {064033} (\bibinfo {year} {2026})},\ \Eprint
  {https://arxiv.org/abs/2509.05479} {arXiv:2509.05479 [gr-qc]} \BibitemShut
  {NoStop}%
\bibitem [{\citenamefont {{Cornish}}(2020)}]{Cornish:2020odn}%
  \BibitemOpen
  \bibfield  {author} {\bibinfo {author} {\bibfnamefont {N.~J.}\ \bibnamefont
  {{Cornish}}},\ }\href {https://doi.org/10.1103/PhysRevD.102.124038}
  {\bibfield  {journal} {\bibinfo  {journal} {\prd}\ }\textbf {\bibinfo
  {volume} {102}},\ \bibinfo {eid} {124038} (\bibinfo {year} {2020})},\ \Eprint
  {https://arxiv.org/abs/2009.00043} {arXiv:2009.00043 [gr-qc]} \BibitemShut
  {NoStop}%
\bibitem [{\citenamefont {{Cornish}}(2025)}]{cornish_nonstationary_wdm}%
  \BibitemOpen
  \bibfield  {author} {\bibinfo {author} {\bibfnamefont {N.~J.}\ \bibnamefont
  {{Cornish}}},\ }\href {https://doi.org/10.48550/arXiv.2511.10632} {\bibfield
  {journal} {\bibinfo  {journal} {arXiv e-prints}\ ,\ \bibinfo {eid}
  {arXiv:2511.10632}} (\bibinfo {year} {2025})},\ \Eprint
  {https://arxiv.org/abs/2511.10632} {arXiv:2511.10632 [gr-qc]} \BibitemShut
  {NoStop}%
\bibitem [{\citenamefont {Cornish}(2020)}]{Cornish_WDM_Transform}%
  \BibitemOpen
  \bibfield  {author} {\bibinfo {author} {\bibfnamefont {N.~J.}\ \bibnamefont
  {Cornish}},\ }\href@noop {} {\bibinfo {title} {{WDM\_Transform}: Codes to
  compute the {WDM} wavelet transform}},\ \bibinfo {howpublished}
  {\url{https://github.com/eXtremeGravityInstitute/WDM_Transform}} (\bibinfo
  {year} {2020}),\ \bibinfo {note} {eXtreme Gravity Institute, Montana State
  University. Implements fast {TaylorT}, {TaylorF}, and sparse {WDM} transforms
  for binary chirp signals in C. Accessed: May 2026}\BibitemShut {NoStop}%
\bibitem [{\citenamefont {Digman}(2022)}]{Digman_WDMWaveletTransforms}%
  \BibitemOpen
  \bibfield  {author} {\bibinfo {author} {\bibfnamefont {M.~C.}\ \bibnamefont
  {Digman}},\ }\href@noop {} {\bibinfo {title} {{WDMWaveletTransforms}: Fast
  forward and inverse {WDM} wavelet transforms in {Python}}},\ \bibinfo
  {howpublished} {\url{https://github.com/XGI-MSU/WDMWaveletTransforms}}
  (\bibinfo {year} {2022}),\ \bibinfo {note} {eXtreme Gravity Institute,
  Montana State University. Produced under NASA LISA Preparatory Science Grant
  80NSSC19K0320. Released v0.0.1, April 2025; GPLv2+ licence. Accessed: May
  2026}\BibitemShut {NoStop}%
\bibitem [{\citenamefont {Moore}(2025)}]{Moore_WDM_GW_wavelets}%
  \BibitemOpen
  \bibfield  {author} {\bibinfo {author} {\bibfnamefont {C.~J.}\ \bibnamefont
  {Moore}},\ }\href@noop {} {\bibinfo {title} {{WDM\_GW\_wavelets}: A fast,
  {JAX}-based {Python} implementation of the {Wilson--Daubechies--Meyer}
  wavelet transform for gravitational wave data}},\ \bibinfo {howpublished}
  {\url{https://cjm96.github.io/WDM_GW_wavelets/}} (\bibinfo {year} {2025}),\
  \bibinfo {note} {gitHub repository:
  \url{https://github.com/cjm96/WDM_GW_wavelets}. Includes time-delay filter
  routines for space-based detector responses. Accessed: May 2026}\BibitemShut
  {NoStop}%
\bibitem [{\citenamefont {{cWB Team}}(2024)}]{cWB_software}%
  \BibitemOpen
  \bibfield  {author} {\bibinfo {author} {\bibnamefont {{cWB Team}}},\ }\href
  {https://doi.org/10.5281/zenodo.15032433} {\bibinfo {title} {coherent
  {WaveBurst} ({cWB}), version {cWB-6.4.6.0}}},\ \bibinfo {howpublished}
  {\url{https://gwburst.gitlab.io/documentation/latest/html/}} (\bibinfo {year}
  {2024}),\ \bibinfo {note} {public release used for the LVK O4b analysis.
  GitLab: \url{https://gitlab.com/gwburst/public/library}. GPLv3 licence.
  Accessed: May 2026}\BibitemShut {NoStop}%
\bibitem [{\citenamefont {Farr}(2025)}]{Farr:2025wdm}%
  \BibitemOpen
  \bibfield  {author} {\bibinfo {author} {\bibfnamefont {W.~M.}\ \bibnamefont
  {Farr}},\ }\href@noop {} {\bibinfo {title} {{WDMWavelets.jl: WDM wavelets in
  Julia}}},\ \bibinfo {howpublished}
  {\url{https://github.com/farr/WDMWavelets.jl}} (\bibinfo {year} {2025}),\
  \bibinfo {note} {gitHub repository}\BibitemShut {NoStop}%
\bibitem [{\citenamefont {Krishnan}\ \emph {et~al.}(2004)\citenamefont
  {Krishnan}, \citenamefont {Sintes}, \citenamefont {Papa}, \citenamefont
  {Schutz}, \citenamefont {Frasca},\ and\ \citenamefont
  {Palomba}}]{Krishnan:2004sv}%
  \BibitemOpen
  \bibfield  {author} {\bibinfo {author} {\bibfnamefont {B.}~\bibnamefont
  {Krishnan}}, \bibinfo {author} {\bibfnamefont {A.~M.}\ \bibnamefont
  {Sintes}}, \bibinfo {author} {\bibfnamefont {M.~A.}\ \bibnamefont {Papa}},
  \bibinfo {author} {\bibfnamefont {B.~F.}\ \bibnamefont {Schutz}}, \bibinfo
  {author} {\bibfnamefont {S.}~\bibnamefont {Frasca}},\ and\ \bibinfo {author}
  {\bibfnamefont {C.}~\bibnamefont {Palomba}},\ }\href
  {https://doi.org/10.1103/PhysRevD.70.082001} {\bibfield  {journal} {\bibinfo
  {journal} {Phys. Rev. D}\ }\textbf {\bibinfo {volume} {70}},\ \bibinfo
  {pages} {082001} (\bibinfo {year} {2004})},\ \Eprint
  {https://arxiv.org/abs/gr-qc/0407001} {arXiv:gr-qc/0407001} \BibitemShut
  {NoStop}%
\bibitem [{\citenamefont {Palomba}\ \emph {et~al.}(2005)\citenamefont
  {Palomba}, \citenamefont {Astone},\ and\ \citenamefont
  {Frasca}}]{Palomba_2005}%
  \BibitemOpen
  \bibfield  {author} {\bibinfo {author} {\bibfnamefont {C.}~\bibnamefont
  {Palomba}}, \bibinfo {author} {\bibfnamefont {P.}~\bibnamefont {Astone}},\
  and\ \bibinfo {author} {\bibfnamefont {S.}~\bibnamefont {Frasca}},\ }\href
  {https://doi.org/10.1088/0264-9381/22/18/S39} {\bibfield  {journal} {\bibinfo
   {journal} {Classical and Quantum Gravity}\ }\textbf {\bibinfo {volume}
  {22}},\ \bibinfo {pages} {S1255} (\bibinfo {year} {2005})}\BibitemShut
  {NoStop}%
\bibitem [{\citenamefont {Savalle}\ \emph {et~al.}(2022)\citenamefont
  {Savalle}, \citenamefont {Gair}, \citenamefont {Speri},\ and\ \citenamefont
  {Babak}}]{Savalle_2022}%
  \BibitemOpen
  \bibfield  {author} {\bibinfo {author} {\bibfnamefont {E.}~\bibnamefont
  {Savalle}}, \bibinfo {author} {\bibfnamefont {J.}~\bibnamefont {Gair}},
  \bibinfo {author} {\bibfnamefont {L.}~\bibnamefont {Speri}},\ and\ \bibinfo
  {author} {\bibfnamefont {S.}~\bibnamefont {Babak}},\ }\href
  {https://doi.org/10.1103/PhysRevD.106.022003} {\bibfield  {journal} {\bibinfo
   {journal} {Phys. Rev. D}\ }\textbf {\bibinfo {volume} {106}},\ \bibinfo
  {pages} {022003} (\bibinfo {year} {2022})}\BibitemShut {NoStop}%
\bibitem [{\citenamefont {Johnson}\ \emph {et~al.}(2026)\citenamefont
  {Johnson}, \citenamefont {Chatziioannou},\ and\ \citenamefont
  {Summers}}]{Johnson:2026rrn}%
  \BibitemOpen
  \bibfield  {author} {\bibinfo {author} {\bibfnamefont {A.}~\bibnamefont
  {Johnson}}, \bibinfo {author} {\bibfnamefont {K.}~\bibnamefont
  {Chatziioannou}},\ and\ \bibinfo {author} {\bibfnamefont {J.}~\bibnamefont
  {Summers}},\ }\href@noop {} {\bibfield  {journal} {\bibinfo  {journal} {-}\ }
  (\bibinfo {year} {2026})},\ \Eprint {https://arxiv.org/abs/2606.21473}
  {arXiv:2606.21473 [gr-qc]} \BibitemShut {NoStop}%
\bibitem [{\citenamefont {Daubechies}\ \emph {et~al.}(1991)\citenamefont
  {Daubechies}, \citenamefont {Jaffard},\ and\ \citenamefont
  {Journe}}]{Daubechies:1991wv}%
  \BibitemOpen
  \bibfield  {author} {\bibinfo {author} {\bibfnamefont {I.}~\bibnamefont
  {Daubechies}}, \bibinfo {author} {\bibfnamefont {S.}~\bibnamefont
  {Jaffard}},\ and\ \bibinfo {author} {\bibfnamefont {J.~L.}\ \bibnamefont
  {Journe}},\ }\href {https://doi.org/10.1137/0522035} {\bibfield  {journal}
  {\bibinfo  {journal} {SIAM J. Math. Anal.}\ }\textbf {\bibinfo {volume}
  {22}},\ \bibinfo {pages} {554} (\bibinfo {year} {1991})}\BibitemShut
  {NoStop}%
\bibitem [{\citenamefont {{Burke}}\ \emph {et~al.}(2025)\citenamefont
  {{Burke}}, \citenamefont {{Marsat}}, \citenamefont {{Gair}},\ and\
  \citenamefont {{Katz}}}]{Burke:2025bun}%
  \BibitemOpen
  \bibfield  {author} {\bibinfo {author} {\bibfnamefont {O.}~\bibnamefont
  {{Burke}}}, \bibinfo {author} {\bibfnamefont {S.}~\bibnamefont {{Marsat}}},
  \bibinfo {author} {\bibfnamefont {J.~R.}\ \bibnamefont {{Gair}}},\ and\
  \bibinfo {author} {\bibfnamefont {M.~L.}\ \bibnamefont {{Katz}}},\ }\href
  {https://doi.org/10.1103/5jr8-k2ss} {\bibfield  {journal} {\bibinfo
  {journal} {\prd}\ }\textbf {\bibinfo {volume} {111}},\ \bibinfo {eid}
  {124053} (\bibinfo {year} {2025})},\ \Eprint
  {https://arxiv.org/abs/2502.17426} {arXiv:2502.17426 [gr-qc]} \BibitemShut
  {NoStop}%
\bibitem [{\citenamefont {Bartolo}\ \emph {et~al.}(2022)\citenamefont {Bartolo}
  \emph {et~al.}}]{LISACosmologyWorkingGroup:2022kbp}%
  \BibitemOpen
  \bibfield  {author} {\bibinfo {author} {\bibfnamefont {N.}~\bibnamefont
  {Bartolo}} \emph {et~al.},\ }\href
  {https://doi.org/10.1088/1475-7516/2022/11/009} {\bibfield  {journal}
  {\bibinfo  {journal} {\jcap}\ }\textbf {\bibinfo {volume} {2022}},\ \bibinfo
  {eid} {009} (\bibinfo {year} {2022})},\ \Eprint
  {https://arxiv.org/abs/2201.08782} {arXiv:2201.08782 [astro-ph.CO]}
  \BibitemShut {NoStop}%
\bibitem [{\citenamefont {Bayle}\ \emph {et~al.}(2025)\citenamefont {Bayle},
  \citenamefont {Le~Jeune},\ and\ \citenamefont {Menu}}]{jaxgb}%
  \BibitemOpen
  \bibfield  {author} {\bibinfo {author} {\bibfnamefont {J.-B.}\ \bibnamefont
  {Bayle}}, \bibinfo {author} {\bibfnamefont {M.}~\bibnamefont {Le~Jeune}},\
  and\ \bibinfo {author} {\bibfnamefont {J.}~\bibnamefont {Menu}},\ }\href@noop
  {} {\bibinfo {title} {{jaxgb: Fast LISA response for Galactic binaries using
  JAX}}},\ \bibinfo {howpublished} {\url{https://pypi.org/project/jaxgb/}}
  (\bibinfo {year} {2025}),\ \bibinfo {note} {version 0.2.1,
  BSD-3-Clause}\BibitemShut {NoStop}%
\bibitem [{\citenamefont {{Phan}}\ \emph {et~al.}(2019)\citenamefont {{Phan}},
  \citenamefont {{Pradhan}},\ and\ \citenamefont
  {{Jankowiak}}}]{phan2019composable}%
  \BibitemOpen
  \bibfield  {author} {\bibinfo {author} {\bibfnamefont {D.}~\bibnamefont
  {{Phan}}}, \bibinfo {author} {\bibfnamefont {N.}~\bibnamefont {{Pradhan}}},\
  and\ \bibinfo {author} {\bibfnamefont {M.}~\bibnamefont {{Jankowiak}}},\
  }\href {https://doi.org/10.48550/arXiv.1912.11554} {\bibfield  {journal}
  {\bibinfo  {journal} {arXiv e-prints}\ ,\ \bibinfo {eid} {arXiv:1912.11554}}
  (\bibinfo {year} {2019})},\ \Eprint {https://arxiv.org/abs/1912.11554}
  {arXiv:1912.11554 [stat.ML]} \BibitemShut {NoStop}%
\bibitem [{\citenamefont {Klimenko}\ \emph {et~al.}(2005)\citenamefont
  {Klimenko}, \citenamefont {Mohanty}, \citenamefont {Rakhmanov},\ and\
  \citenamefont {Mitselmakher}}]{Klimenko:2005xv}%
  \BibitemOpen
  \bibfield  {author} {\bibinfo {author} {\bibfnamefont {S.}~\bibnamefont
  {Klimenko}}, \bibinfo {author} {\bibfnamefont {S.}~\bibnamefont {Mohanty}},
  \bibinfo {author} {\bibfnamefont {M.}~\bibnamefont {Rakhmanov}},\ and\
  \bibinfo {author} {\bibfnamefont {G.}~\bibnamefont {Mitselmakher}},\ }\href
  {https://doi.org/10.1103/PhysRevD.72.122002} {\bibfield  {journal} {\bibinfo
  {journal} {Phys. Rev. D}\ }\textbf {\bibinfo {volume} {72}},\ \bibinfo
  {pages} {122002} (\bibinfo {year} {2005})},\ \Eprint
  {https://arxiv.org/abs/gr-qc/0508068} {arXiv:gr-qc/0508068} \BibitemShut
  {NoStop}%
\bibitem [{\citenamefont {Klimenko}\ \emph {et~al.}(2016)\citenamefont
  {Klimenko} \emph {et~al.}}]{Klimenko:2015ypf}%
  \BibitemOpen
  \bibfield  {author} {\bibinfo {author} {\bibfnamefont {S.}~\bibnamefont
  {Klimenko}} \emph {et~al.},\ }\href
  {https://doi.org/10.1103/PhysRevD.93.042004} {\bibfield  {journal} {\bibinfo
  {journal} {Phys. Rev. D}\ }\textbf {\bibinfo {volume} {93}},\ \bibinfo
  {pages} {042004} (\bibinfo {year} {2016})},\ \Eprint
  {https://arxiv.org/abs/1511.05999} {arXiv:1511.05999 [gr-qc]} \BibitemShut
  {NoStop}%
\bibitem [{\citenamefont {Vajpeyi}\ \emph {et~al.}(2026)\citenamefont
  {Vajpeyi}, \citenamefont {Mentasti}, \citenamefont {Baghi}, \citenamefont
  {Burke},\ and\ \citenamefont {Speri}}]{wdm_transform_zenodo}%
  \BibitemOpen
  \bibfield  {author} {\bibinfo {author} {\bibfnamefont {A.}~\bibnamefont
  {Vajpeyi}}, \bibinfo {author} {\bibfnamefont {G.}~\bibnamefont {Mentasti}},
  \bibinfo {author} {\bibfnamefont {Q.}~\bibnamefont {Baghi}}, \bibinfo
  {author} {\bibfnamefont {O.}~\bibnamefont {Burke}},\ and\ \bibinfo {author}
  {\bibfnamefont {L.}~\bibnamefont {Speri}},\ }\href
  {https://doi.org/10.5281/zenodo.20740881} {\bibinfo {title}
  {{pywavelet/wdm\_transform: v0.05}}} (\bibinfo {year} {2026})\BibitemShut
  {NoStop}%
\end{thebibliography}%

\end{document}